\documentclass[fleqn, usenatbib]{mnras}

\usepackage{newtxtext, newtxmath}

\usepackage[T1]{fontenc}

\DeclareRobustCommand{\VAN}[3]{#2}
\let\VANthebibliography\thebibliography
\def\thebibliography{\DeclareRobustCommand{\VAN}[3]{##3}\VANthebibliography}

\usepackage{graphicx}	
\usepackage{amsmath}

\graphicspath{{./}{figures/}}

\usepackage{subcaption}

\usepackage{longtable}

\usepackage{lscape} 

\usepackage{tabularx}

\usepackage{newtxtext, newtxmath}

\newcommand{\Msun}{\ensuremath{M_{\odot}}}

\title[GDR of M17]{The Gas-to-Dust Ratio Investigation in the Massive Star-Forming region M17}

\author[Qi Zhao et al.]{
Qi Zhao,$^{1}$
Zhibo Jiang,$^{2}$
Chao-Jian Wu,$^{1}$\thanks{E-mail: chjwu@bao.ac.cn}
Jie Zheng,$^{1}$
Cheng Cheng,$^{1}$
YiNan Zhu,$^{1}$
Hong Wu,$^{1}$
\\
$^{1}$CAS Key Laboratory of Optical Astronomy, National Astronomical Observatories, Chinese Academy of Sciences, Beijing 100101, China\\
$^{2}$Purple Mountain Observatory, Chinese Academy of Sciences, 10 Yuanhua Road, 210023 Nanjing, China\\
}

\date{Accepted XXX. Received YYY; in original form ZZZ}

\pubyear{\the\year{}}

\begin{document}
\label{firstpage}
\pagerange{\pageref{firstpage}--\pageref{lastpage}}

\maketitle

\begin{abstract}
M17 is a well-known massive star-forming region, and its Gas-to-Dust Ratio (GDR) may vary significantly compared to the other areas. 
The mass of gas can be traced by the ${\rm CO}$ emission observed in the \emph{Milky Way Imaging Scroll Painting (MWISP) project}.
The dust mass can be traced by analyzing the interstellar extinction magnitude obtained from the \emph{United Kingdom Infrared Telescope (UKIRT)}. 
We computed the ratio ${W({\rm CO})/A_V}$: for ${A_V \le }$ 10 mag,  ${{ W(^{12}{\rm CO})/ A_V}= (6.27 \pm 0.19)}$  ${\mathrm{{K \cdot km/s} \cdot mag^{-1}}}$ and ${{ W(^{13}{\rm CO})/ A_V} = (0.75 \pm 0.72)}$  ${ \mathrm{{K \cdot km/s} \cdot mag^{-1}}}$; whereas for ${{A_V} \ge 10}$ mag,  ${{ W(^{12}{\rm CO})/ A_V} = (15.8 \pm 0.06) }$  ${\mathrm{{K \cdot km/s} \cdot mag^{-1}}}$ and ${{ W(^{13}{\rm CO})/ A_V} = (3.11 \pm 0.25)}$  ${ \mathrm{{K \cdot km/s} \cdot mag^{-1}}}$.
Then, we converted the ${W({\rm CO})/A_V}$ into ${N(\rm H)/A_V}$.
Using the WD01 model, we derived the GDR: for ${A_V \le }$ 10 mag, the GDRs were  ${118 \pm 9}$ for ${^{12}{\rm CO}}$ and ${83 \pm 62}$ for ${^{13}{\rm CO}}$, comparable to those of the Milky Way; however, for ${A_V \ge }$ 10 mag, the GDRs increased significantly to  ${296 \pm 3}$ for ${^{12}{\rm CO}}$ and ${387 \pm 40}$ for ${^{13}{\rm CO}}$, approximately three times higher than those of the Milky Way.
In the discussion, we compared the results of this work with previous studies and provided a detailed discussion of the influence of massive stars and other factors on GDR.

\end{abstract}




\section{Introduction}\label{sec: intro}

Studies have demonstrated that stars are formed within molecular clouds consisting of atomic gas, molecular gas, and dust \citep{1987ARA&A..25...23S}. The molecular cloud holds a pivotal position in the process of star formation.
Delving into the mechanisms behind star formation is paramount for understanding molecular clouds' origin, development, and physical characteristics.
Consequently, monitoring the dust and gas components is a fundamental tool in mapping the configuration and dispersal of molecular clouds within star-forming regions \citep{2015MNRAS.448.2187C}.

In terms of gas mass, molecular hydrogen (${\rm H_{2}}$) constitutes the primary component of molecular clouds (MCs) within the interstellar medium (ISM). However, due to the absence of a permanent dipole moment and the corresponding dipolar rotational transitions in the radio band, ${\rm H_{2}}$ radiates inefficiently in the cold, dense regions of the molecular ISM \citep{2013ARA&A..51..207B}.
Fortunately, carbon monoxide (${\rm CO}$) exhibits a strong interaction with ${\rm H_{2}}$ and is the second most abundant molecule in the ISM. Consequently, ${\rm CO}$ is extensively utilized to trace molecular gas, as its emission is quickly excited in the molecular ISM environment. Furthermore, the ${{\rm CO}}$ (${J = 1 \to 0}$) transition at 2.6 mm (or 115 GHz) is conveniently observable from the ground \citep{1991ARA&A..29..195C} \citep{2001ApJ...547..792D} \citep{2015ARA&A..53..583H} \citep{2015MNRAS.448.2187C}.
Therefore, the ${\rm CO}$ integrated intensity (${W({\rm CO})}$) is commonly employed as a proxy for representing the column density of molecular hydrogen ${N(\rm H_{2})}$.

By measuring the extinction of background starlight caused by clouds at visible and near-infrared wavelengths, we can ascertain the dust mass and delineate the comprehensive distribution of dust within these clouds.
Following convention, we represented the dust column density in terms of the extinction magnitude in the V band (${A_{V}}$).
With the reddening law \citep{1985ApJ...288..618R}, we can convert the mean colour excess of each star to an equivalent visual mean extinction magnitude.

The GDR value of the Milky Way is widely accepted to be around 100. 
\citet{1978ppim.book.....S} suggested a range of GDR values between 20 and 700 for star-forming regions, whereas \citet{1991ARA&A..29..581Y} derived an average GDR of 600 for molecular clouds and \citet{1995A&A...300..493L} calculated a GDR of 450 for L1688. These results indicate that GDR values are higher than Milky Way's. 
\citet{2015A&A...578A.131L} utilized ${\rm N_{2}H^{+}}$ as a gas tracer to investigate the star-forming region ${\rho}$ ${Oph}$ ${A}$, deriving an average GDR value of 88. 
The GDR value in the central regions around the core typically hovers around 100. They attribute this discrepancy to the influence of dust grain size on the GDR. Additionally, in the dense core region, both ${\rm N_{2}H^{+}}$ ions and ${\rm CO}$ molecules will freeze onto the surface of dust grains, consequently leading to a lower estimated gas mass \citep{2015A&A...578A.131L}.

The M17 complex harbours one of the most massive molecular clouds in the Milky Way \citep{1974ApJ...189L..35L}, making it one of the brightest ${\rm H II}$ regions in the sky. 
Observations have revealed that the M17 cluster contains a large population of young stellar objects (YSOs) and massive stars (5\Msun ${\sim}$ 20\Msun ). As a result, the GDR in M17 might differ from that in other regions significantly \citep{1997ApJ...489..698H}. 
So, M17 was selected as the focus of this work to investigate the impact of massive star formation on GDR.

In this paper, we explored the correlation between the integrated intensity of ${\rm CO}$ and the extinction magnitude in the V band, denoted as ${{W({\rm CO})}/{A_V}}$, and conducted a series of discussions. Section \ref{sec: data} covers data processing, Section \ref{sec: result} details our findings, and in Section \ref{sec: discussion}, we analyze and discuss our results. Finally, Section \ref{sec: summary} presents an overview of our findings.

\section{data} 
\label{sec: data}

The GDR can be expressed by the ratio of gas column density to dust column density, given by: 

\begin{equation}
  GDR=\frac{N(gas)}{N(dust)}\sim\frac{W({\rm CO})}{A_V}
    \label{equ: GDR}
\end{equation}
since the primary component within molecular clouds is ${N(\rm H_{2})}$, we represent the gas column density of the molecular cloud by the column density of molecular hydrogen (${N(\rm H_{2})}$). Furthermore, due to the coupling between ${\rm H_{2}}$ and ${\rm CO}$, we use $W({\rm CO})$ to represent the gas column density of the molecular cloud. Since $A_{V}$ is directly proportional to the dust column density, we use $A_V$ to represent the dust column density. $W({\rm CO})$ is derived from the \emph{Milky Way Imaging Scroll Painting project (MWISP)}, while $A_{V}$ is derived from near-infrared ${J}$-, ${H}$- and ${K}$- band photometry of M17, with the near-infrared imaging sourced from the PI project U/05A/J2 at the \emph{United Kingdom Infrared Telescope (UKIRT)}.

\subsection{MWISP}

The \emph{MWISP project} entails a northern Galactic plane CO survey conducted with the \emph{13.7 m millimetre-wavelength telescope} located in Delingha, China (referred to as the DLH telescope). Led by the Purple Mountain Observatory (PMO), the MWISP project receives comprehensive support from the staff members at Delingha.
The DLH telescope is located near Delingha, Qinghai Province, China, at 37°22'4'' N, 97°33'6'' E. At around 3200 meters, Delingha benefits from dry and stable atmospheric conditions, rendering it an ideal site for millimetre astronomy \citep{2019ApJS..240....9S}.

The \emph{MWISP project} is a large-scale, unbiased, and high-sensitivity triple ${\rm CO}$ isotope line survey targeting the northern Galactic plane (with Galactic longitudes ranging from $l=-10^{\circ}$ to $+250^{\circ}$ and Galactic latitudes $|b| \lesssim 5^{\circ}$) utilizing a 13.7-meter single-dish telescope.
The sky coverage of the \emph{MWISP project} is divided into 10941 cells, with each cell comprising a ${30^{\prime} \times 30^{\prime}}$ 3D image.
The \emph{MWISP} CO survey offers three key advantages:   
1) It conducts large-scale ${\rm CO}$ mapping with high spatial dynamics;
2)It provides an unbiased CO survey with high sensitivity;
3)It enables simultaneous observations of ${^{12}{\rm CO}}$, ${^{13}{\rm CO}}$, and ${\rm C^{18}O}$ (${J = 1 \to 0}$) line transitions.
With a spatial resolution of approximately ${\sim 50^{\prime \prime}}$ and a grid spacing of ${30^{\prime \prime}}$, the full-sampling \emph{MWISP} survey furnishes a rich ${\rm CO}$ dataset for regions with ${b \ge 1^{\circ}}$, which are less covered by other ${\rm CO}$ surveys. 
The typical rms noise level of ${\sim 0.5}$ K for ${^{12}{\rm CO}}$ and ${\sim 0.3}$ K for ${^{13}{\rm CO}}$.
The \emph{MWISP} ${\rm CO}$ survey provides a valuable opportunity to investigate molecular cloud properties \citep{2019ApJS..240....9S}.

\subsection{UKIRT}

The \emph{UKIRT} has a diameter of 3.8 meters, and it is situated at the Mauna Kea Observatory on Hawaii Island, USA, at 155°28'14"W, 19°49'21"N, an altitude of approximately 4194 meters.
\emph{UKIRT} has advanced instruments, including sophisticated cameras and filters. 
One such instrument is the Wide Field Camera (WFCAM), which operates in the ${J}$, ${H}$ and ${K}$ near-infrared bands \citep{2007A&A...467..777C}.
It offers significantly greater depth compared to the \emph{Two-Micron All-Sky Survey} (2MASS; \citet{2006AJ....131.1163S}). For M17, the limiting magnitudes are ${J=22.5}$ mag, ${H=20.5}$ mag, and ${K=20.0}$ mag.
In contrast, the limiting magnitudes derived from 2MASS are ${J=18.5}$ mag, ${H=17.3}$ mag, and ${K=16.3}$ mag.
Therefore, we can observe a more significant number of background stars, facilitating the calculation of the ${A_{V}}$ for the entire molecular cloud.

\subsection{Data reduction}

The cell size of the \emph{MWISP} CO survey is ${30^{\prime} \times 30^{\prime}}$, while the region encompassing M17 spans $1^{\circ} \times 1^{\circ}$. Therefore, to construct a 3D image, we must merge these cells. Additionally, given the ${61^{\circ}.877}$ angle between the Galactic Coordinate System and the Equatorial Coordinate System, we  determined that the jointing range should be ${l=14^{\circ}.43 \sim 15^{\circ}.67}$ and ${b=-1^{\circ}.31 \sim -0^{\circ}.05}$.
Next, we utilized the CLASS software to visualize the jointed 3D images. Through analysis, we identified the integral interval as ranging from ${10}$ km/s to ${27}$ km/s for ${^{12}{\rm CO}}$ and from ${10}$ km/s to ${25}$  km/s for ${^{13}{\rm CO}}$.
Thirdly, we generated integrated intensity maps for ${^{12}{\rm CO}}$ and ${^{13}{\rm CO}}$ within the specified integral intervals.
Fourthly, we rotated the maps by an angle of ${61^{\circ}.877}$ and transformed the Galactic Coordinate System into the Equatorial Coordinate System.
Fifthly, we cropped the integrated intensity maps based on the coordinates of M17, ${RA=18^{\rm h}18^{\rm m}30^{\rm s} \sim 18^{\rm h}22^{\rm m}23^{\rm s}}$, and ${DEC=-16^{\circ}37^{\prime}50^{\prime \prime} \sim -15^{\circ}43^{\prime}50^{\prime \prime}}$.

\begin{figure*}
  \centering
  \includegraphics[width=\textwidth]{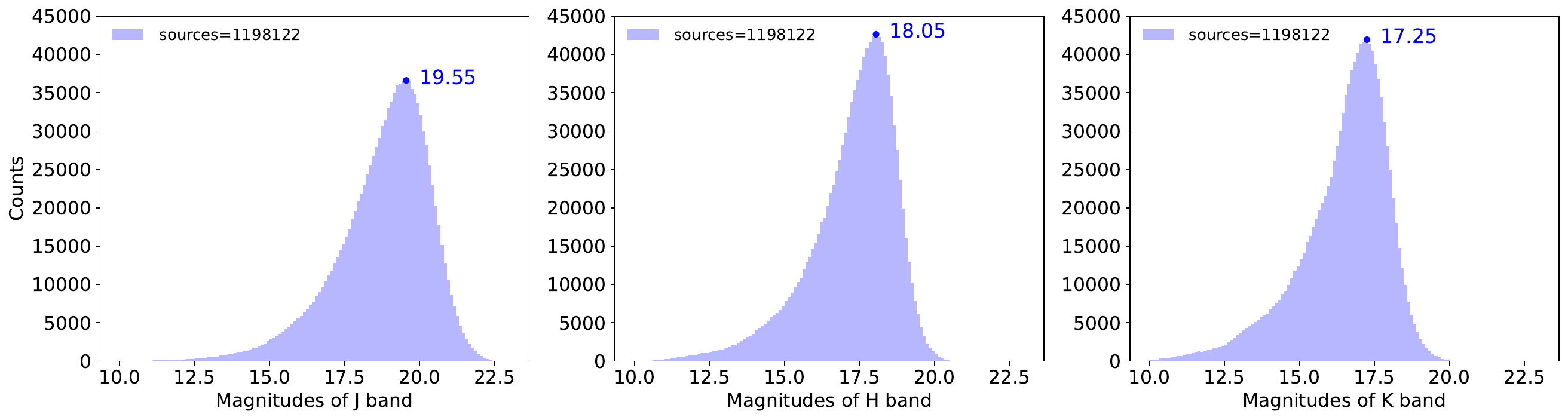}
  \caption{ Histograms of the apparent magnitude in the ${J}$-, ${H}$- and ${K}$-bands of M17. We observed approximately 1.20 million stars. In the ${J}$-band, the apparent magnitude is concentrated in 19.55 mag; in the ${H}$-band, the apparent magnitude is concentrated in 18.05 mag; while in the ${K}$-band, the apparent magnitude is concentrated in 17.25 mag.}
  \label{fig: apparent luminosity function}
\end{figure*}

Due to the dense population of stars in M17, we opted for the Point Source Function (PSF) photometry method to compute the photometric magnitudes with \textsc{iraf}/\textsc{daophot}.
We set the threshold to 4$\sigma$ and used scripts such as \textsc{daofind}, \textsc{phot}, \textsc{psf}, \textsc{seepsf}, and \textsc{allstar} for PSF photometry. Specifically, we initially selected as many isolated stars when creating the PSF model with the \textsc{psf} script. Through repeated testing and removal of unsuitable isolated stars, we obtained a PSF model that can accurately reflect the image. We performed PSF photometry using this model with the \textsc{allstar} script. After receiving the residual map, we re-evaluated the PSF model. If it was not satisfactory, we returned to creating the PSF model, removed unsuitable isolated stars, modified the PSF model, and re-evaluated the residual map. We repeated this process until the brightness variations of the residual map were no longer significant. Then, we performed a second PSF photometry on the residual map. 
Next, we made the zero magnitude calibration with the standard star from the \emph{UKIRT} dataset via Python. 
Firstly, generating distribution plots, we conducted a cross-match between the \emph{UKIRT} standard catalogue and the PSF photometric star catalogue. Secondly, we selected stars from the distribution plot, ranging from the peak to one magnitude brighter than the peak, with an error margin of less than 0.1 mag. Thirdly, we implemented a 3$\sigma$ filter and performed Gaussian fitting to derive the mean values and standard deviation. Lastly, we calibrated zero star magnitude by applying the mean value to the PSF photometric star catalogue.
Ultimately, we identified approximately 1.20 million stars across the ${J}$-, ${H}$- and ${K}$-bands. 
In the ${J}$-band, the complete mag is 19.55; in the ${H}$-band, the complete mag is 18.05; while in the ${K}$-band, the complete mag is 17.25.
Figure \ref{fig: apparent luminosity function} depicts histograms illustrating the apparent magnitude distributions of stars within M17.

\begin{figure}
  \includegraphics[width=\columnwidth]{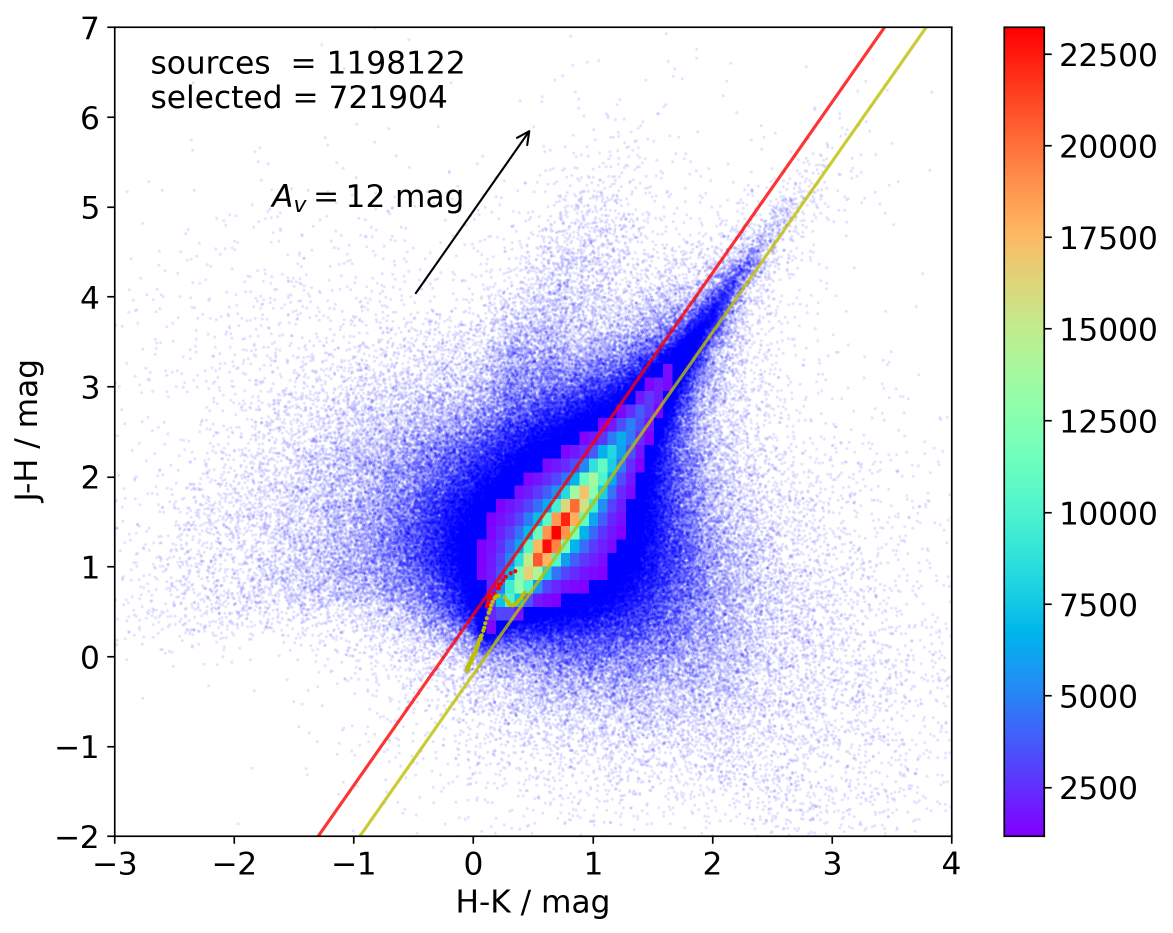}
  \caption{ Colour-colour diagram created by the PSF photometry star catalogue. The horizontal axis represents the colour index of ${(H-K)}$, and the vertical axis represents the colour index of ${(J-H)}$. The red dots represent giant star trajectories, while the yellow dots represent main-sequence dwarf trajectories. We constructed an envelope of the giant branch trajectories and the main-sequence dwarf trajectory at the second turn, employing a slope of 1.90 and accounting for photometric errors of 0.1 magnitude.  The region between the two envelope lines is the reddening band, and the stars of the reddening band will be used to calculate ${A_{V}}$.}
  \label{fig: colour-colour}
\end{figure}

Using the PSF photometric star catalogue, we generated a colour-colour diagram illustrated in Figure \ref{fig: colour-colour} to make the reddening band calculate ${A_{V}}$.
Interstellar reddening causes a star to shift along the reddening vector toward the upper-right direction on the colour-colour diagram. Therefore, in the presence of only interstellar reddening, the coordinates of a star on the colour-colour diagram will move toward the upper-right corner along this vector. This shifted band is called the reddening band. Stars within the reddening band are often used for extinction calculations.
Drawing upon astronomy and astrophysics constants, we derived trajectories for main sequence dwarf and giant stars. Assuming a slope value of 1.90 for the reddening vector \citep{1998A&A...329..161C}, we constructed an envelope of the giant branch trajectories as the blue limit of the reddening band, and the main-sequence dwarf trajectory (after the second turn point) as the red limit of the reddening band. Furthermore, we extended the reddening band by 0.1 magnitude in both directions to account for photometric errors.
After filtering the colour-colour diagram, we ultimately identified 723000 stars.

The magnitude difference between any two bands of the same star is called the colour index, which indicates the star's surface temperature. A star's intrinsic colour index represents its inherent colour corresponding to its specific spectrum, effectively reflecting the energy distribution of the star's continuous spectrum within a certain band.
Due to interstellar extinction, the intrinsic colour index of a star becomes reddened. Therefore, based on the observed magnitudes, we calculate the observed colour index of each star. The relationship between the colour excess and the observed colour index and intrinsic colour index is as follows \citep{1997IAUS..170...47A}: 

\begin{equation}
  {E}({H}-{K})=({H}-{K})_{\text {observed }}-({H}-{K})_{\text {intrinsic }}
\end{equation}
${E}({H}-{K})$ is the colour excess, $({H}-{K})_{\text {observed }}$ the observed colour index, and $({H}-{K})_{\text {intrinsic }}$ the intrinsic colour index. The relationship between extinction and colour excess is as follows \citep{1999ApJ...512..250L}: 

\begin{equation}
  A_{\lambda}=r_{\lambda}^{{H}}, { }^{{K}} E(H-K)=r_{\lambda}^{{H}}, { }^{{K}} [({H}-{K})_{\text {observed }}-({H}-{K})_{\text {intrinsic }}]
\end{equation}
$A_{\lambda}$ represents the extinction in the $\lambda$ band, while $r_{\lambda}^{{H}}, { }^{{K}}$ denotes the conversion coefficient used to translate the colour excess of the H and K bands into extinction in the $\lambda$ band.

According to \citet{2017ApJ...838...80C}, ${{({H}-{K})_{\text {intrinsic }}}=0.2}$, ${r_{V}^{{H}}, { }^{{K}}=15}$, and the relationship between ${A_V}$ and ${H}$- and ${K}$-bands are as follows: 

\begin{equation}
  \label{equ: Av}
  {A}_{V} = 15 \times [({H}_{o b s} - {K}_{o b s}) - 0.2]
\end{equation}
Hence, we can determine ${A_{V}}$ by computing the ${(H_{obs}-K_{obs})}$ value for each background star as it traverses through the molecular cloud. 
The typical uncertainties are 0.9 mag in ${{A}_{V}}$, or about 0.06 mag in ${{E}({H}-{K})}$.

\begin{figure}
	\includegraphics[width=\columnwidth]{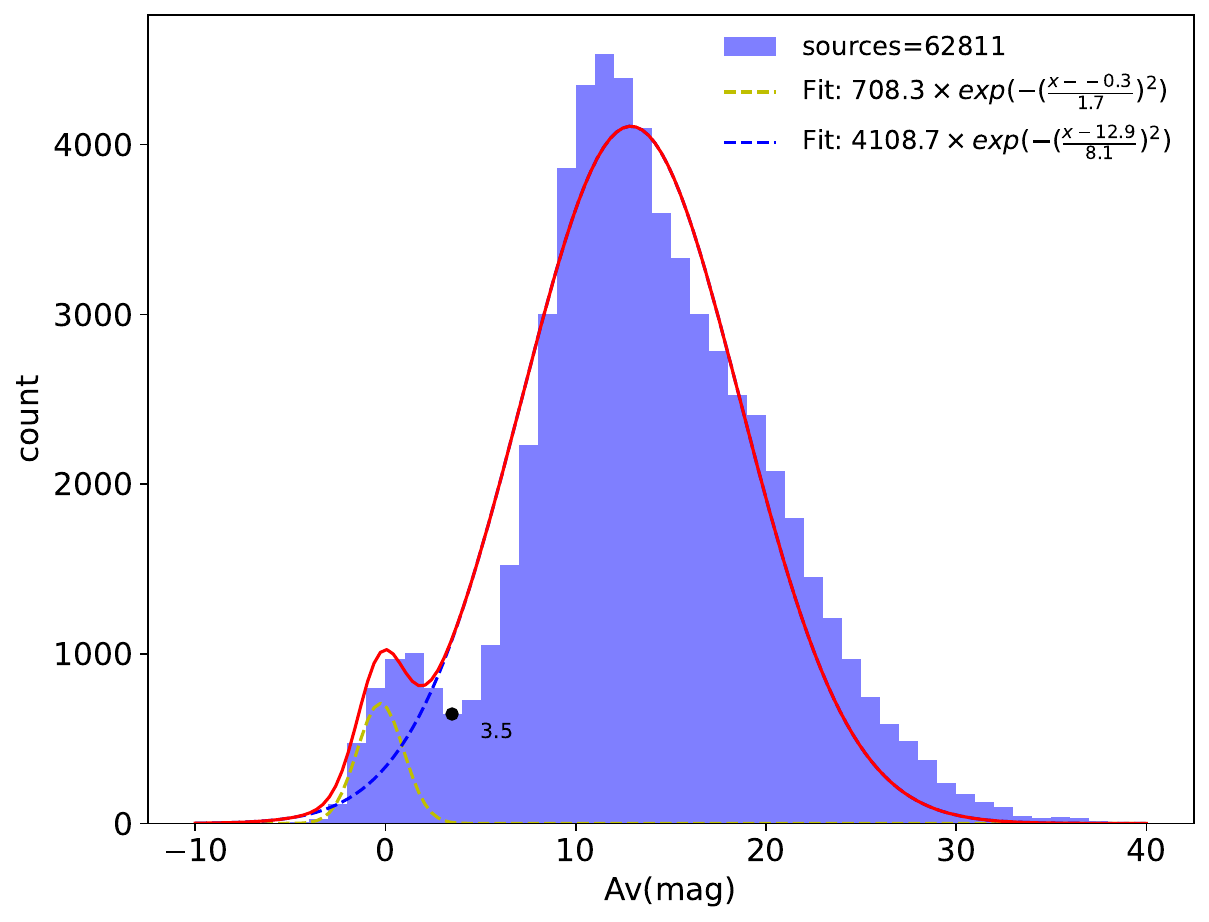}
  \caption{ Histogram of the original ${A_{V}}$ value of the high extinction pixels (${A_V>12}$ mag). We identified the high extinction by each pixel's average ${A_V>12}$ mag. Subsequently, we performed a double Gaussian fit on the original ${A_{V}}$ mag of each star in these pixels. Additionally, we found that the lowest point between the two peaks is 3.5 mag. Therefore, we consider stars with an extinction value of less than 3.5 mag foreground stars and should be excluded.}
  \label{fig: double-Guassin}
\end{figure}

Subsequently, we proceeded to eliminate potential foreground stars. By partitioning the area into ${30^{\prime \prime} \times 30^{\prime \prime}}$ pixel cells, we calculated the average ${A_{V}}$ value of stars within each cell and generated an extinction map.
Following this, assuming cells with an average extinction magnitude ${A_V>12}$ mag denoted high extinction areas, we performed a double-Gaussian fitting line using the original ${A_{V}}$ values of stars within these cells. In Figure \ref{fig: double-Guassin}, we determined the lowest point value between the two peaks to be 3.5 mag. Consequently, stars with ${A_{V}}$ values below 3.5 mag were classified as foreground stars \citep{2011A&A...536A..48K}, and then we removed them.

Finally, we generated the extinction map using a two-dimensional Gaussian function and computed the weighted average value.
To calculate the extinction value of each pixel and derive the weighted average, we employed the 2D Gaussian function. The formula for the 2D Gaussian function is as follows: 

\begin{equation}
\label{equ: 2D Guassian}
  f(x, y)=A \times e^{-\frac{\left(x-x_{0}\right)^{2}+\left(y-y_{0}\right)^{2}}{2 \sigma^{2}}}
\end{equation}
Here, A represents the star's extinction magnitude ${A_V}$, while x and y correspond to the star's Right Ascension (RA) and Declination (DEC) coordinates, respectively. Additionally, ${x_0}$ and ${y_0}$ denote the central coordinates of each pixel.
The relationship between the full width at half maximum (FWHM) and $\sigma$ for a normal distribution is given by ${\sigma = FWHM/2.35}$. In this work, the FWHM represents the resolution we intentionally set at 90". With this established, we proceed to generate the extinction map of M17.

\section{result} 
\label{sec: result}

\subsection{{\rm CO} integrated intensity maps and extinction map}

\begin{figure}
	\includegraphics[width=\columnwidth]{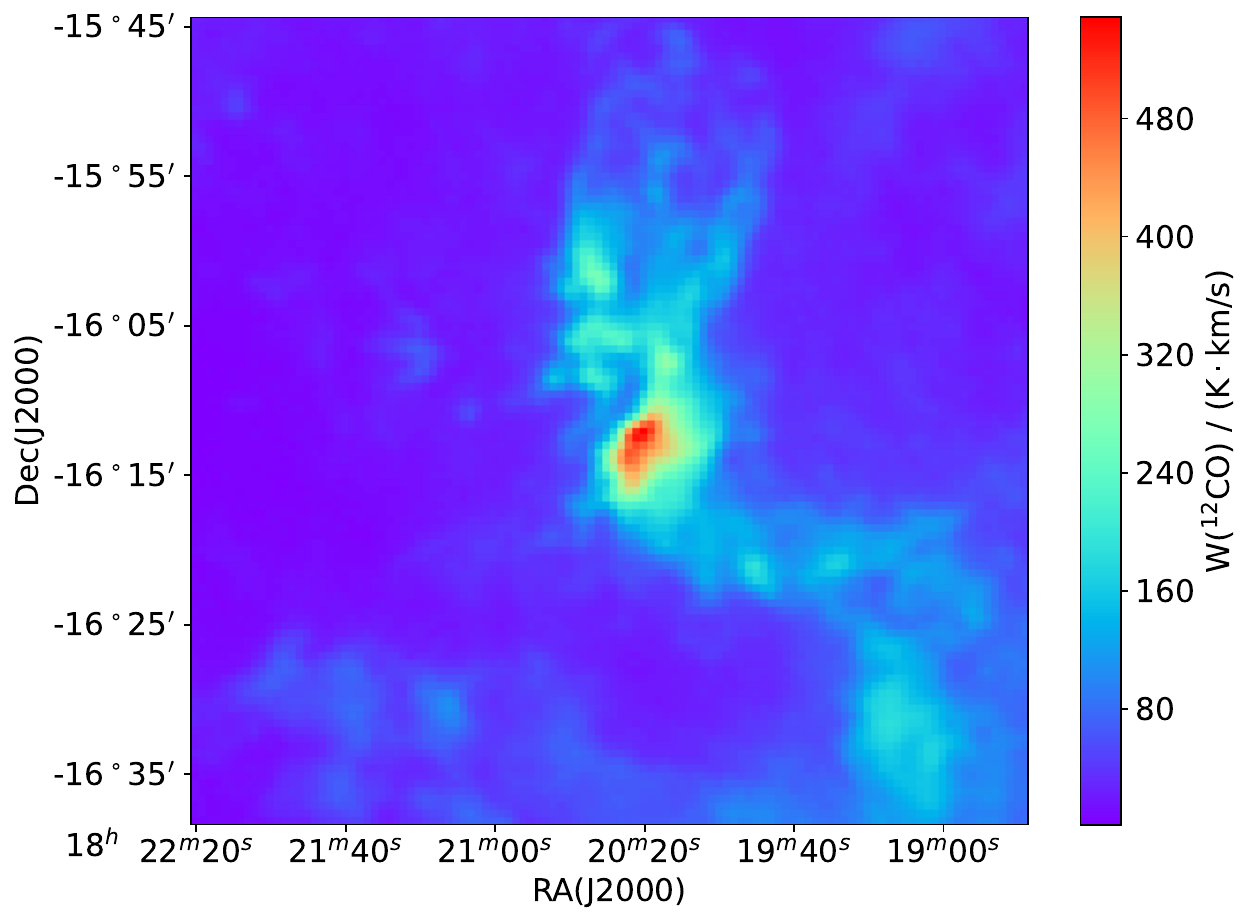}
  \caption{ ${^{12}{\rm CO}}$ integrated intensity map of M17 in the equatorial coordinate system. The pixel size of the map is ${30^{\prime \prime} \times 30^{\prime \prime}}$. The unit is ${\mathrm{K \cdot km/s}}$.}
  \label{fig: W12}
\end{figure}

\begin{figure}
	\includegraphics[width=\columnwidth]{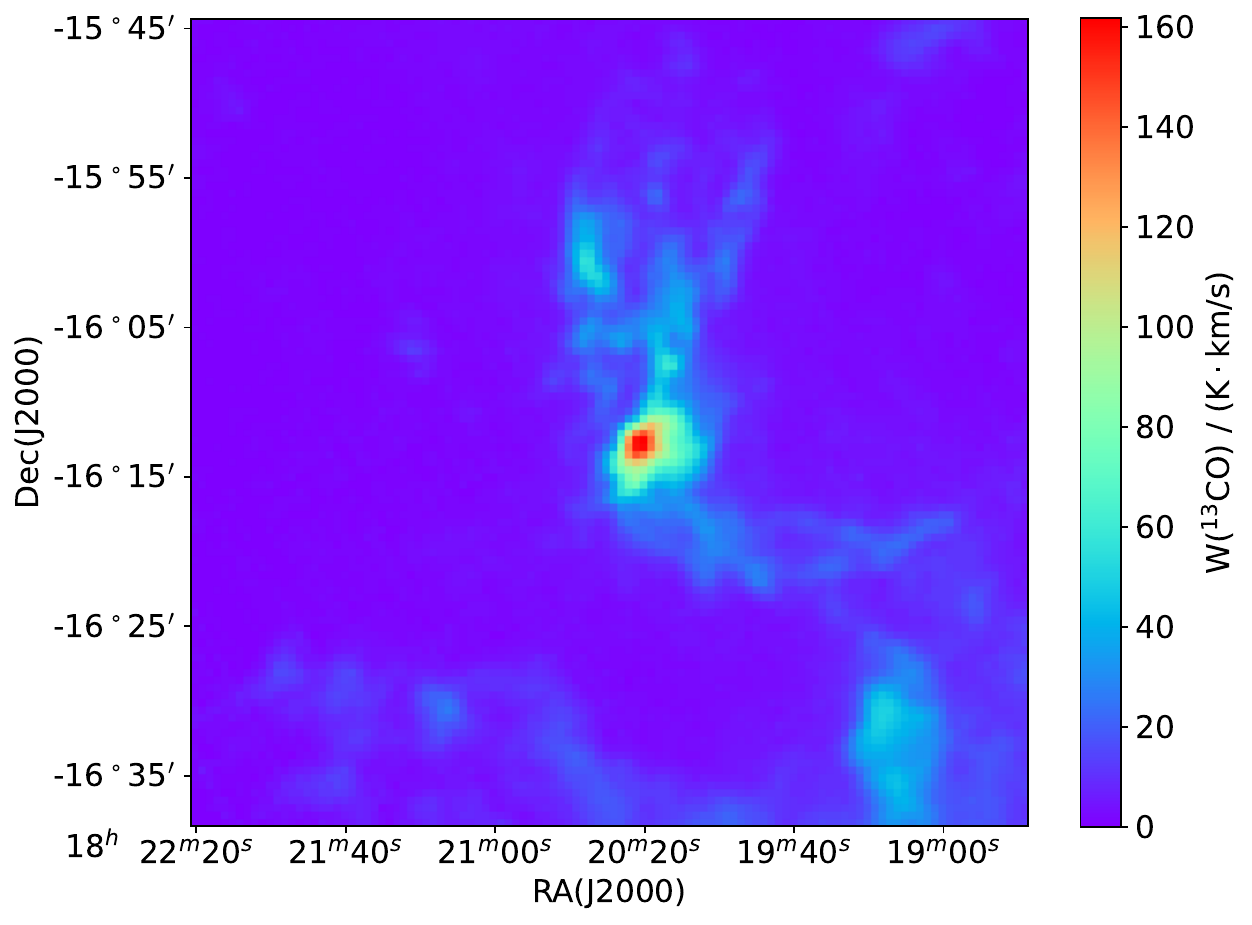}
  \caption{ ${^{13}{\rm CO}}$ integrated intensity map of M17 in the equatorial coordinate system. The pixel size of the map is ${30^{\prime \prime} \times 30^{\prime \prime}}$. The unit is ${\mathrm{K \cdot km/s}}$.}
  \label{fig: W13}
\end{figure}

\begin{figure}
	\includegraphics[width=\columnwidth]{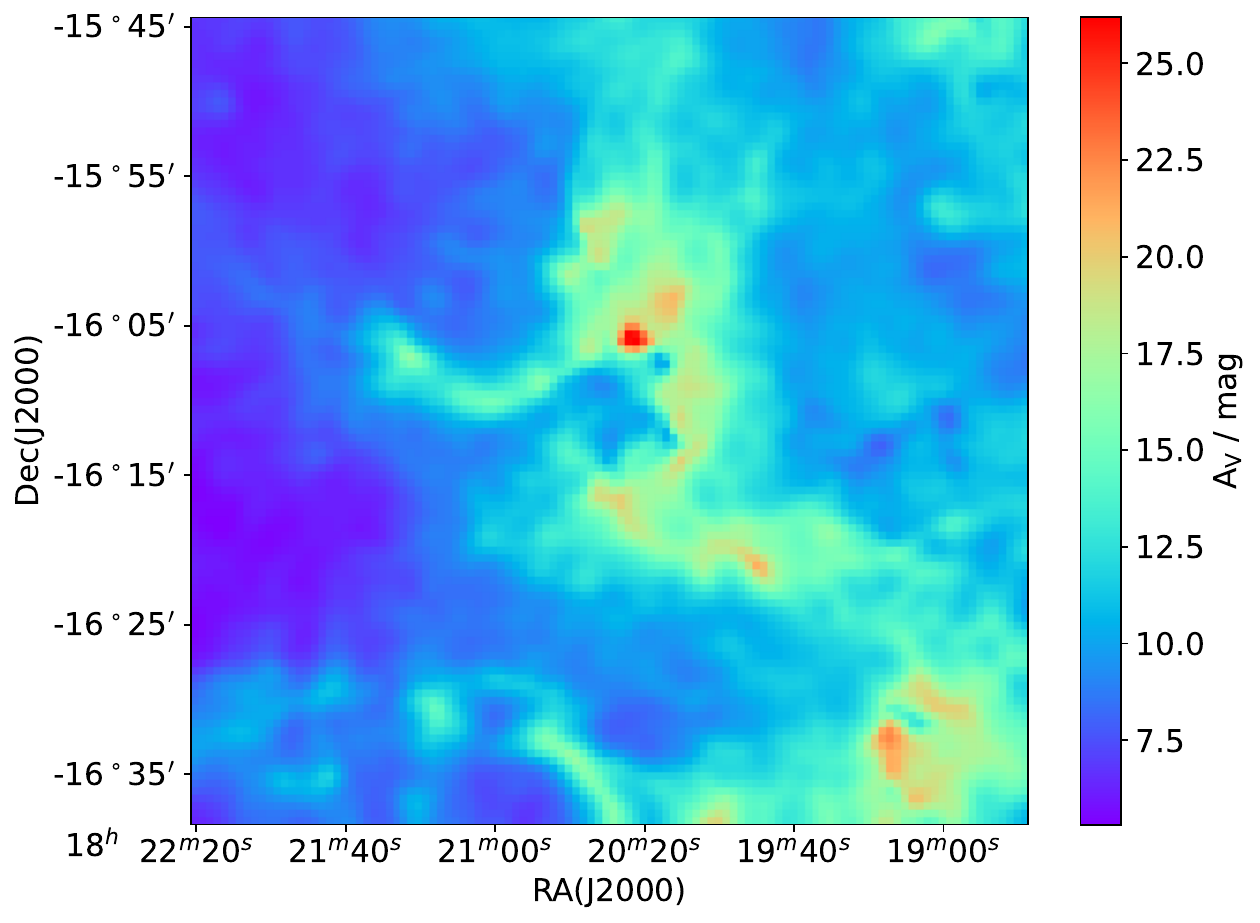}
  \caption{ Extinction map of M17 in the equatorial coordinate system. The rainbow colourmap ranges from ${A_V=}$ 5 mag (purple) to ${A_V=}$ 26 mag (red). We utilized ${30^{\prime \prime} \times 30^{\prime \prime}}$ as the pixel cell size, yielding a resolution of ${90^{\prime \prime}}$.  The unit is mag.}
  \label{fig: extinction}
\end{figure}

The ${\rm CO}$ integrated intensity maps are depicted in Figure \ref{fig: W12} and \ref{fig: W13}. And the extinction map of the M17 region is presented in Figure \ref{fig: extinction}.

\citet{2008MNRAS.391..136L} utilized \emph{UKIDSS-GPS} data from 2005 to compute the extinction map for the ${J}$-, ${H}$- and ${K}$-bands of M17 by PSF photomety with \textsc{iraf}/\textsc{daophot}.
However, they solely utilized the \textsc{daofind} script and conducted zero magnitude calibration using the WSA catalogue. Furthermore, they ultimately detected 0.788 million stars. 
In comparison, we  detected approximately 1.2 million more stars and attained lower photometric errors.
For example, in the ${J}$-band, when the magnitude is greater than 15, our photometric errors are lower than theirs at the same magnitude.
They calculated ${A_{V}}$, averaging stars within a ${15^{\prime \prime} \times 15^{\prime \prime}}$ pixel cell and subsequently smoothed the FWHM with 1.5-pixel cells.
However, we utilize the two-dimensional Gaussian function to craft the extinction map, employing a weighted average method to calculate the value of each pixel, thus achieving greater accuracy in this work.
Although our pixel cell size is ${30^{\prime \prime} \times 30^{\prime \prime}}$, matching the \emph{MWISP} data and larger than that used by \citet{2008MNRAS.391..136L}.

\subsection{\texorpdfstring{${W({\rm CO})/A_V}$}{} of M17}

\begin{figure}
	\includegraphics[width=\columnwidth]{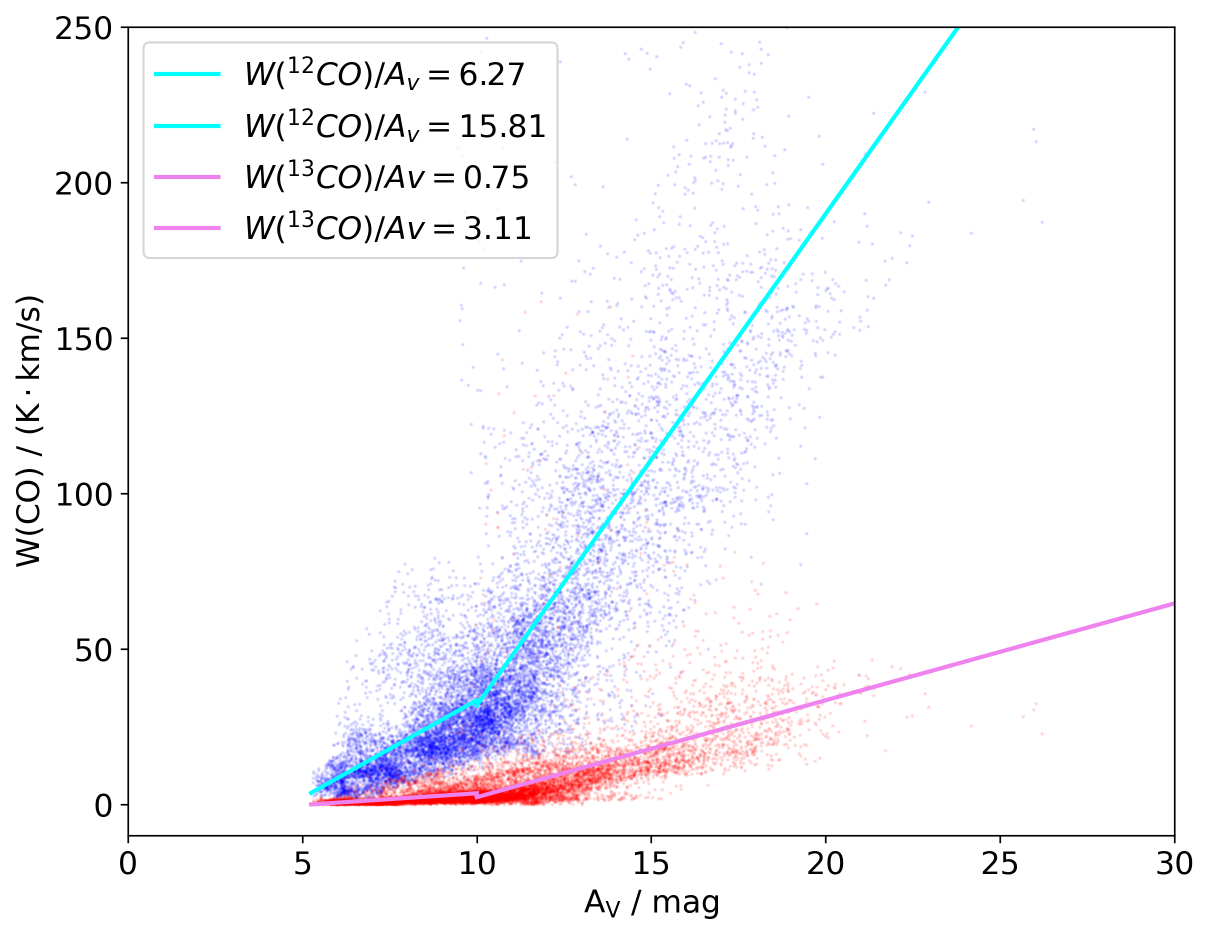}
  \caption{ ${W({\rm CO})/A_V}$ ratio derived from ${^{12}{\rm CO}}$ integrated intensity map and ${^{13}{\rm CO}}$ integrated intensity map. The blue dots represent the scatter plot of ${W(^{12}{\rm CO})/A_V}$, with the blue-green line representing the fitted line of ${W(^{12}{\rm CO})/A_V}$. Similarly, the red dots depict the scatter plot of ${W(^{13}{\rm CO})/A_V}$, and the pink line represents the fitted line of ${W(^{13 }{\rm CO})/A_V}$.  The unit is  $\mathrm{{K \cdot km/s} \cdot mag^{-1}}$.}
  \label{fig: W1213Av}
\end{figure}

Using ${A_V}$ as the horizontal coordinate and ${W({\rm CO})}$ as the vertical coordinate, we generated a scatter plot for each pixel. We determined the slope in Figure \ref{fig: W1213Av}, where the slope value represents ${W({\rm CO})/A_V}$.

There are two linear relationships present.
In \citet{2018ChAA..42..213L}, they opted to downscale the resolution due to data originating from various telescopes with differing resolutions. Consequently, the maxima of the data were altered; for instance, the maximum extinction ${A_V}$ was reduced to 9 mag. To facilitate a comparison between this work and theirs, we established ${A_V}= 10$ mag as the cutoff point.
Based on the cutoff point, we applied a segmented fit. 
When ${{A_V} \le 10}$ mag, ${{ W(^{12}{\rm CO})/ A_V}= (6.27 \pm 0.19)}$ ${\mathrm{{K \cdot km/s} \cdot mag^{-1}}}$ derived from ${^{12}{\rm CO}}$, while ${{ W(^{13}{\rm CO})/ A_V} = (0.75 \pm 0.72)}$ ${\mathrm{{K \cdot km/s} \cdot mag^{-1}}}$ derived from ${^{13}{\rm CO}}$. Conversely, when ${{A_V} \ge 10}$ mag, ${{ W(^{12}{\rm CO})/ A_V} = (15.81 \pm 0.06)}$ ${ \mathrm{{K \cdot km/s} \cdot mag^{-1}}}$ derived from ${^{12}{\rm CO}}$, and ${{ W(^{13}{\rm CO})/ A_V} = (3.11 \pm 0.25)}$ ${\mathrm{{K \cdot km/s} \cdot mag^{-1}}}$ derived from ${^{13}{\rm CO}}$.

\subsection{\texorpdfstring{${N(\rm H)/A_V}$}{} of M17}

Since the GDR represents the ratio of gas mass to dust mass, we need to convert ${W({\rm CO})}$ into ${N(\rm H)}$.
For ${^{12}{\rm CO}}$, according to \citet{MolecularAstrophysics}, there exists a correlation between ${W({^{12}{\rm CO}})}$ and ${\rm H_{2}}$, with the conversion coefficient denoted as ${X_{\rm CO}}$. Assuming that ${X_{\rm CO}}$ is linked to ${N(\rm H_{2})}$ and the surface brightness emitted by ${^{12}{\rm CO}}$,  \citet{1986ApJ...309..326D} established the relationship between ${W({^{12}{\rm CO}}}$ ${J = 1 \to 0)}$ (unit: $\mathrm{K \cdot km/s}$) and ${N({\rm H_{2}})}$ (unit: $\mathrm{cm^{-2}}$): 

\begin{equation}
  \label{equ: XCO}
  {N}({H}_{2})={X}_{{\rm CO}} \times {W}({ }^{12} {\rm CO})
\end{equation}
where ${W(^{12}{\rm CO})=\int T_{A} dv}$ represents the integral intensity of the ${^{12}{\rm CO}}$ spectral line. 
 {We adopt} ${X_{\rm CO}=2 \times 10^{20}}$  ${\mathrm{cm^{-2} \cdot {(K \cdot km/s)}^{-1}}}$, and its uncertainty is about ${\pm 30\%}$ \citep{2013ARA&A..51..207B}.

\begin{figure}
	\includegraphics[width=\columnwidth]{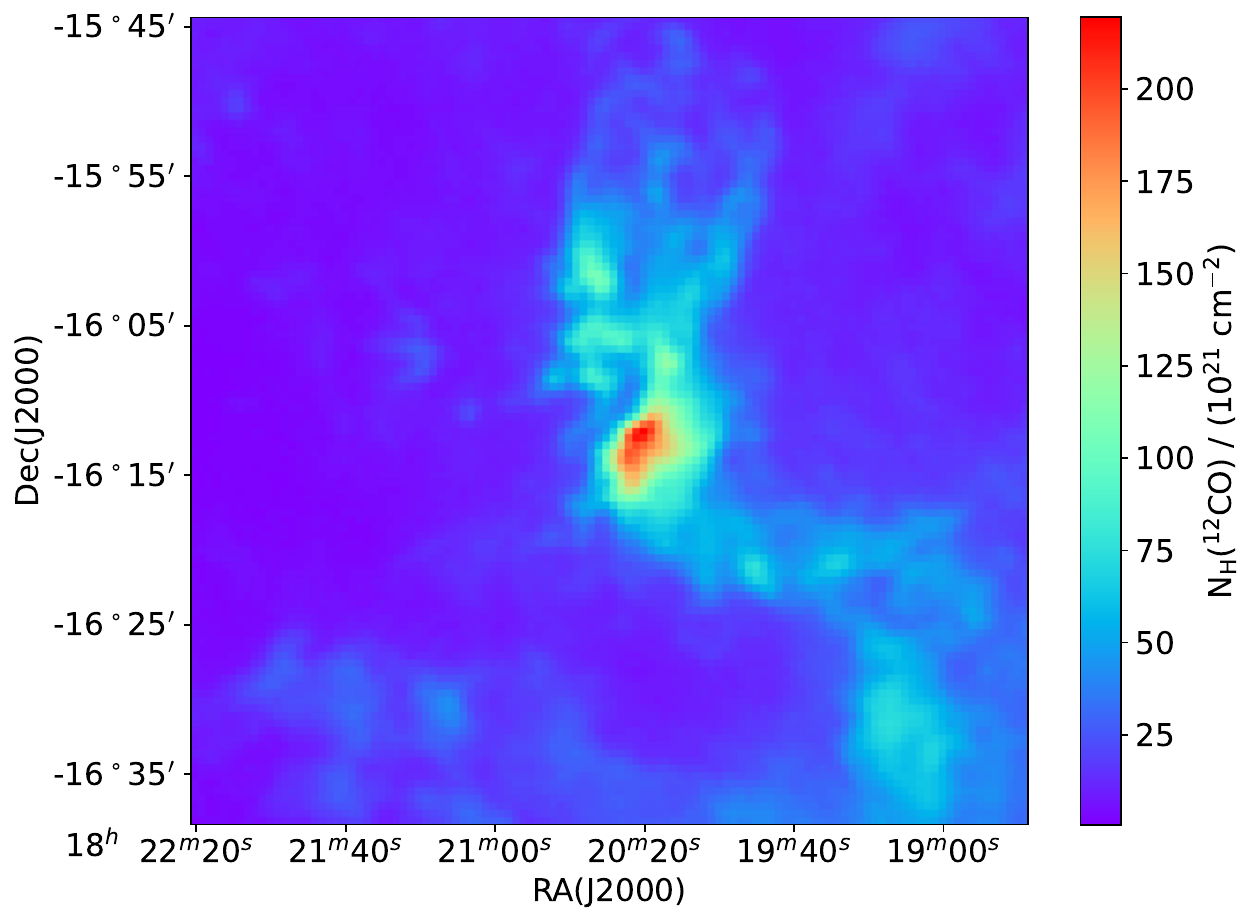}
  \caption{ The column density of hydrogen (${N({\rm H})}$) map derived from ${^{12}{\rm CO}}$ of M17 in the equatorial coordinate system. The pixel size of the map is ${30^{\prime \prime} \times 30^{\prime \prime}}$. The unit is ${10^{21}}$  $\mathrm{cm^{-2}}$.}
  \label{fig: NH12CO}
\end{figure}

According to \citet{2018ChAA..42..213L}, we convert ${N({\rm H})}$ to ${N({\rm H_{2}})}$ with:

\begin{equation}
  {N({\rm H})=N({\rm H_{2}}) \times 2}
  \label{eq: N(H)}
\end{equation}
allowing us to compute the column density of hydrogen (${N({\rm H})}$) derived from ${^{12}{\rm CO}}$,  and the map is depicted in Figure \ref{fig: NH12CO}.

Furthermore, for ${^{13}{\rm CO}}$, according to \citet{MolecularAstrophysics}, we computed the column density of hydrogen element with the following formulas.
Under the assumption of local thermodynamic equilibrium (LTE), and considering that the column density of ${^{13}{\rm CO}}$ can be inferred from the spectral peaks of ${^{12}{\rm CO}}$ and ${^{13}{\rm CO}}$, we derive the ${\rm CO}$ excitation temperature ${T_{ex}}$ from the observed spectral line radiation temperature (${T^*_{R}(^{12}{\rm CO})}$) of ${^{12}{\rm CO}}$ ${J = 1 \to 0}$.
In most instances, ${^{12}{\rm CO}}$ molecules are primarily influenced by collisions, thus ${T_{ex}}$ can be approximated as the kinetic temperature of the molecular clouds: 

\begin{equation}
  \mathit{T}_{\mathit{ex}}=5.532\left[\ln \left[1+\frac{5.532}{\left(\mathit{~T}_{\mathit{R}}^{*}\left({ }^{12} C O\right)+0.819\right)}\right]\right]^{-1}
\end{equation}
Here, ${T^*_{R}(^{12}{\rm CO})}$denotes 
the corrected peak of the ${^{12}{\rm CO}}$ spectrum.

In general, ${^{13}{\rm CO}}$ ${J = 1 \to 0}$ is considered optically thin, meaning $\tau (^{13}{\rm CO})<1$. Assuming that within the same molecular cloud, $T_{ex}(^{12}{\rm CO})=T_{ex}(^{13}{\rm CO})$, we can employ the following formulas from \citet{MolecularAstrophysics}: 

\begin{equation}
  J_{1}\left(T_{e x}\right)=\frac{1}{e^{\frac{5.29}{T_{e x}}-1}}
\end{equation}

\begin{equation}
  \tau\left({ }^{13} C O\right)=-\ln \left[1-\frac{\mathit{T}_{\mathit{R}}^{*}\left({ }^{13} C O\right)}{5.29\left[J_{1}\left(T_{e x}\right)-0.164\right]}\right]
\end{equation}

\begin{equation}
  \mathit{N}\left({ }^{13} \mathit{{\rm CO}}\right)=2.42 \times 10^{14} \times \frac{T_{e x} \tau\left({ }^{13} \mathit{{\rm CO}}\right) \Delta v\left({ }^{13} \mathit{{\rm CO}}\right)}{1-e^{\frac{-5.29}{T_{e x}}}} \mathrm{{cm}^{-2}}
\end{equation}
Here, $\Delta v(^{13}{\rm CO})$ represents the Full Width at Half Maximum (FWHM) width of the ${J = 1 \to 0}$ spectral line of ${^{13}{\rm CO}}$, and ${T^*_{R}(^{13}{\rm CO})}$ denotes the corrected peak of the ${^{13}{\rm CO}}$ spectrum.

\begin{figure}
	\includegraphics[width=\columnwidth]{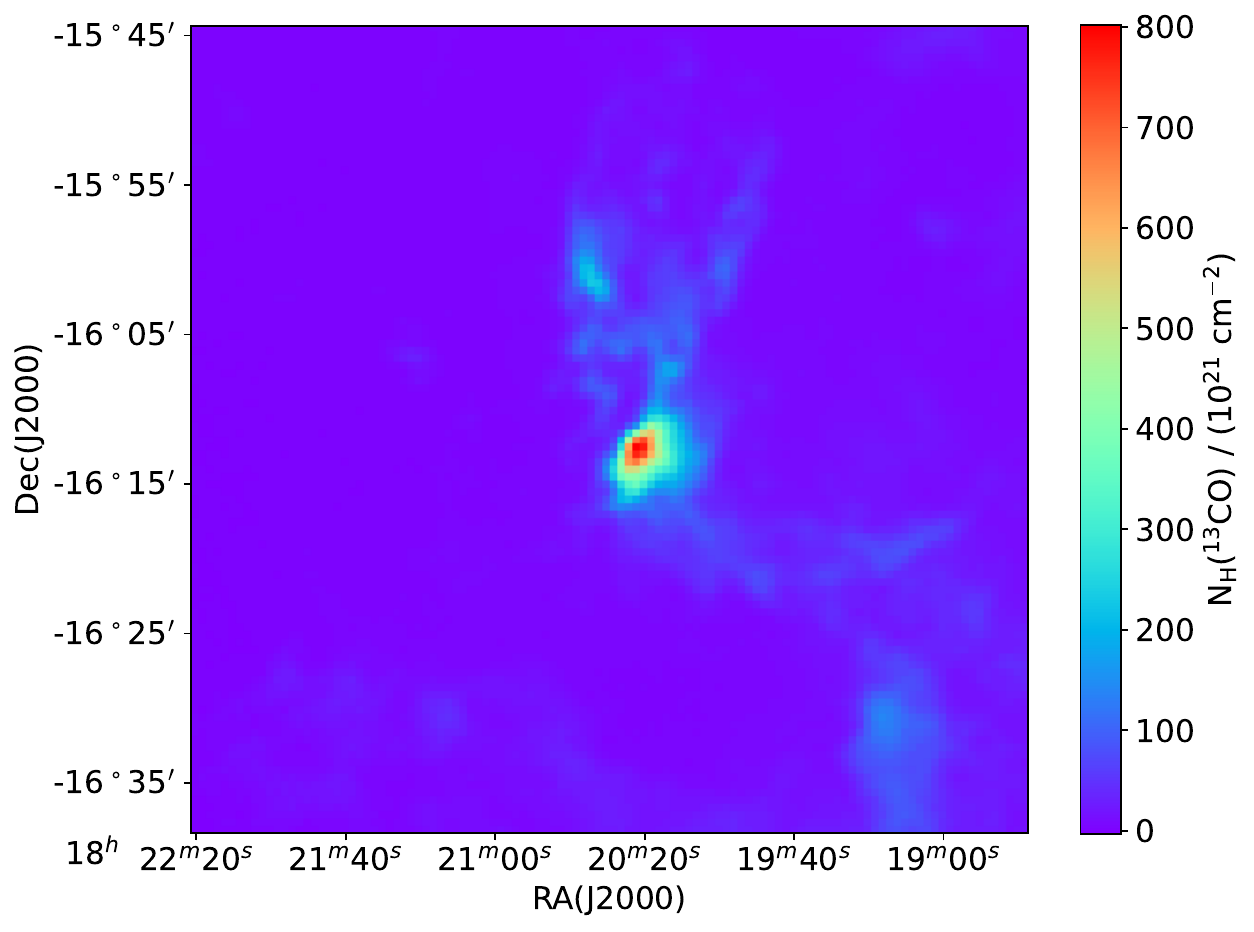}
  \caption{ The column density of hydrogen (${N({\rm H})}$) map derived from ${^{13}{\rm CO}}$ of M17 in the equatorial coordinate system.  The pixel size of the map is ${30^{\prime \prime} \times 30^{\prime \prime}}$. The unit is ${10^{21}}$  $\mathrm{cm^{-2}}$.}
  \label{fig: NH13CO}
\end{figure}

Assuming the relative abundance of ${^{13}{\rm CO}}$: 

\begin{equation}
  N({H_{2}}) / N({^{13}{\rm CO}})=7.5 \times 10^{5}
\end{equation}
we can compute the column density of ${N({\rm H_{2}})}$. Then, utilizing Equation \ref{eq: N(H)}, we can determine ${N(\rm H)}$ derived from ${^{13}{\rm CO}}$,  and the map is depicted in Figure \ref{fig: NH13CO}. 

\begin{figure}
	\includegraphics[width=\columnwidth]{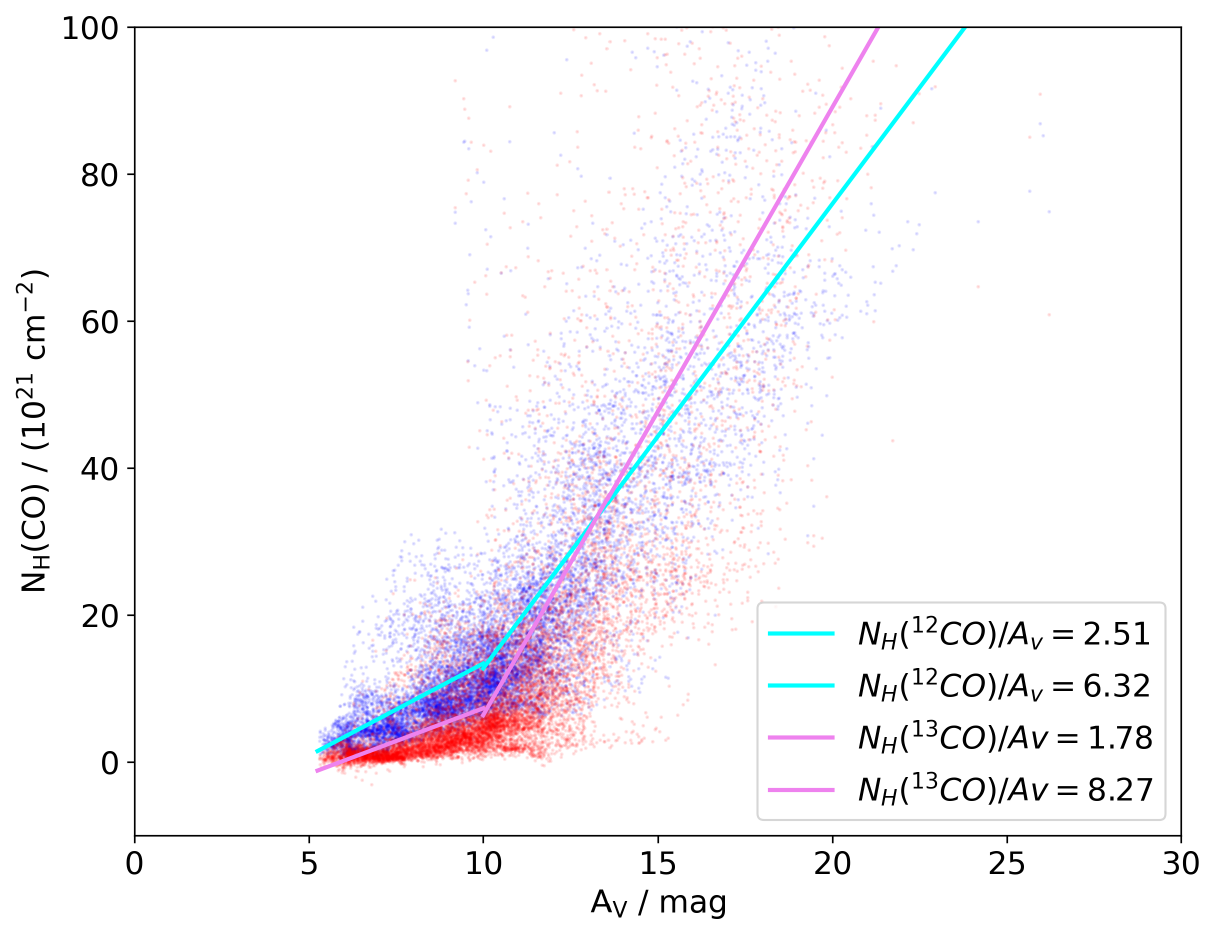}
  \caption{ ${N(\rm H)/A_V}$ ratio derived from ${^{12}{\rm CO}}$ and ${^{13}{\rm CO}}$. The blue dots represent the scatter plot of ${\mathit{N}_{\mathrm{H}}\left({ }^{12} \mathit{\rm CO}\right) / \mathit{A}_{\mathit{V}}}$ derived from ${^{12}{\rm CO}}$, with the blue-green line indicating the fitted line of ${\mathit{N}_{\mathrm{H}}\left({ }^{12} \mathit{\rm CO}\right) / \mathit{A}_{\mathit{V}}}$ derived from ${^{12}{\rm CO}}$. Similarly, the red dots indicate the scatter plot of ${\mathit{N}_{\mathrm{H}}\left({ }^{13} \mathit{\rm CO}\right) / \mathit{A}_{\mathit{V}}}$ derived from ${^{13}{\rm CO}}$, and the pink line represents the fitted line of ${\mathit{N}_{\mathrm{H}}\left({ }^{13} \mathit{\rm CO}\right) / \mathit{A}_{\mathit{V}}}$. The fitting is segmented with ${A_V = }$ 10 mag, when ${A_V \le }$ 10 mag, $\mathit{N}_{\mathrm{H}}\left({ }^{12} \mathit{\rm CO}\right) / \mathit{A}_{\mathit{V}}>\mathit{N}_{\mathrm{H}}\left({ }^{13} \mathit{\rm CO}\right) / \mathit{A}_{\mathit{V}}$; when ${A_V \ge }$ 10 mag, $\mathit{N}_{\mathrm{H}}\left({ }^{12} \mathit{\rm CO}\right) / \mathit{A}_{\mathit{V}}<\mathit{N}_{\mathrm{H}}\left({ }^{13} \mathit{\rm CO}\right) / \mathit{A}_{\mathit{V}}$.  The unit is ${10^{21}}$  $\mathrm{ cm^{-2} \cdot mag^{-1}}$.}
  \label{fig: GDR1213}
\end{figure}

With the ${N(\rm H)}$ maps and the ${A_V}$ map, we can calculate the ${N(\rm H)/A_V}$ ratio.
And the Figure \ref{fig: GDR1213} shows the ${N(\rm H)/A_V}$ values derived from ${^{12}{\rm CO}}$ versus ${^{13}{\rm CO}}$. 
It can be seen that when ${{A_V} \le 10}$ mag, ${N(\rm H)/A_V= (2.51 \pm 0.19)\times 10^{21}}$  ${ \mathrm{cm^{-2} \cdot mag^{-1}}} $ derived from ${^{12}{\rm CO}}$, while ${{N(\rm H)/A_V}= (1.78 \pm 1.33)\times 10^{21} }$  ${\mathrm{cm^{-2} \cdot mag^{-1} }}$ derived from ${^{13}{\rm CO}}$. 
Conversely, when ${{A_V} \ge 10}$ mag, ${N(\rm H)/A_V= (6.32 \pm 0.06) \times 10^{21} }$  ${\mathrm{cm^{-2} \cdot mag^{-1} }}$ derived from ${^{12}{\rm CO}}$, and ${{N(\rm H)/A_V} = (8.27 \pm 0.85) \times 10^{21} }$  ${\mathrm{cm^{-2} \cdot mag^{-1}}}$ derived from ${^{13}{\rm CO}}$.

\subsection{ GDR of M17 } 

The main components of interstellar dust are silicates and graphite, so the simplest dust model consists of only one type of dust, either graphite or silicate.
However, observations of the interstellar extinction curve in the Milky Way suggest that interstellar dust is a blend of both components. Consequently, \citet{2001ApJ...548..296W} derived the WD01 dust model.
And according to \citet{2018ChAA..42..213L}, we can convert the value of ${N(\rm H)/A_V}$ into GDR. For example, the conversion coefficient of silicate can be calculated by the following formula: 
\begin{equation}
  \left(M_{\text {gas }} / M_{\text {dust }}\right) /\left(N(\mathrm{H}) / A_{\mathrm{V}}\right)=25.7 \times 10^{-21} \mathrm{~cm}^{2} \cdot \mathrm{mag}
  \label{eq: GDR}
\end{equation}
Similarly we can calculate the conversion coefficients of graphite and WD01 model to be ${200.2 \times 10^{-21}}$  ${\mathrm{~cm}^{2} \cdot \mathrm{mag}}$ and ${46.8 \times 10^{-21}}$  ${\mathrm{~cm}^{2} \cdot \mathrm{mag}}$, respectively.

For silicate, when ${{A_V} \le }$ 10 mag, the GDR derived from ${^{12}{\rm CO}}$ is {${65 \pm 5}$}, and that from ${^{13}{\rm CO}}$ is {${46 \pm 34}$}, and when ${{A_V} \ge }$ 10 mag, the GDR derived from ${^{12}{\rm CO}}$ is {${162 \pm 2}$}, and that from {${^{13}{\rm CO}}$} is ${213 \pm 22}$. 
For graphite, when ${{A_V} \le }$ 10 mag, the GDR derived from ${^{12}{\rm CO}}$ is {${502 \pm 38}$}, and that from ${^{13}{\rm CO}}$ is {${356 \pm 266}$}, and when ${{A_V} \ge }$ 10 mag, the GDR derived from ${^{12}{\rm CO}}$ is {${1265 \pm 12}$}, and that from ${^{13}{\rm CO}}$ is {${1656 \pm 170}$.}
For WD01 model, when ${{A_V} \le }$ 10 mag, the GDR derived from ${^{12}{\rm CO}}$ is {${118 \pm 9}$}, and that from ${^{13}{\rm CO}}$ is {${83 \pm 62}$}, and when ${{A_V} \ge }$ 10 mag, the GDR derived from ${^{12}{\rm CO}}$ is {${296 \pm 3}$}, and that from ${^{13}{\rm CO}}$ is {${387 \pm 40}$}. Table \ref{tab: Comparison model} lists the comparison.

\begin{table*}
	\centering
	\caption{Comparison of GDR according to different dust models.} 
	\label{tab: Comparison model}
    \begin{tabular}{llllllllll}
    \hline
    \hline
    & \multicolumn{2}{c}{{GDR (${^{12}{\rm CO}}$) \ (This work)}}  & \multicolumn{2}{c}{{GDR (${^{13}{\rm CO}}$) \ (This work)} }   & { GDR \citep{2018ChAA..42..213L} } \\
    & ${A_V \le}$ 10 mag & ${A_V \ge}$ 10 mag & ${A_V \le}$ 10 mag & ${A_V \ge}$ 10 mag & Orion  \\
    \hline
    silicates  &  ${65 \pm 5}$  &  ${162 \pm 2}$   &  ${46 \pm 34}$  &  ${213 \pm 22}$ & 64 \\
    graphite   &  ${502 \pm 38}$   &  ${1265 \pm 12}$    &  ${356 \pm 266}$   &  ${1656 \pm 170}$   & 498 \\
    WD01    &  ${118 \pm 9}$  &  ${296 \pm 3}$   &  ${83 \pm 62}$   &  ${387 \pm 40}$  & 117 \\ 
    \hline
  \end{tabular}
  \label{tab: GDR}
\end{table*}

The GDR derived from different dust compositions can indeed vary significantly. Given that the Milky Way dust comprises a blend of silicates and graphite, the WD01 dust model, which aligns well with the Milky Way extinction curve, is suitable for GDR calculations. When ${{A_V} \le }$ 10 mag, the GDR remains consistent with the typical value for the Milky Way, around 100. However, for ${{A_V} \ge }$ 10 mag, the GDR derived from ${^{12}{\rm CO}}$ and ${^{13}{\rm CO}}$ is approximately three times larger than the commonly observed values. 


Figure \ref{fig: massive stars on GDR of 12CO} and Figure \ref{fig: massive stars on GDR of 13CO} show the GDR maps derived from ${^{12}{\rm CO}}$ and ${^{13}{\rm CO}}$, respectively, and the WD01 dust model was applied when converting from ${N(\rm H)/A_V}$ to GDR.
The GDR maps obtained from both methods exhibit a similar overall distribution.
In regions with higher column density (or extinction), the GDR obtained from ${^{13}{\rm CO}}$ is higher than ${^{12}{\rm CO}}$ due to its thinner optical depth compared to ${^{12}{\rm CO}}$.
Additionally, in denser regions of the molecular cloud, the density of ${^{13}{\rm CO}}$ is also higher compared to the sparse areas, allowing ${^{13}{\rm CO}}$ to more accurately reflect the in the core areas of the molecular cloud.
However, in regions with lower column density (or extinction), the GDR derived from ${^{12}{\rm CO}}$ is slightly larger. Additionally, this situation is more widespread across the entire region because the beam filling factor of ${^{12}{\rm CO}}$ is larger than that of ${^{13}{\rm CO}}$, allowing ${^{12}{\rm CO}}$ to probe a more extensive spatial range. Thus, ${^{12}{\rm CO}}$ better reflects the ${\rm H_{2}}$ in the outer regions of the cloud core.

\section{Discussion} 
\label{sec: discussion}

\subsection{Comparison with previous work} 
\label{subsubsec: comparison}

Since the dust model remains constant, we compared the ${N(\rm H)/A_V}$ values.
Due to the significant differences of the ${N(\rm H)/A_V}$ values in the high-extinction region of M17 compared to previous results, we  concentrated on comparing the results from the low-extinction region with previous studies. 
Since molecular clouds are relatively sparse in areas with low extinction, and ${^{12}{\rm CO}}$ is more abundant compared to ${^{13}{\rm CO}}$, and ${^{12}{\rm CO}}$ has a more extended spatial distribution. A larger beam filling factor, ${^{12}{\rm CO}}$ is more accurately to reflect ${N(\rm H)}$ under these conditions. Therefore, we chose ${N(\rm H)/A_V}$ of ${^{12}{\rm CO}}$ to make the subsequent comparisons.
Upon comparison,  we found that our result  ${N(\rm H)/A_V= (2.51 \pm 0.19)\times 10^{21}}$  ${\mathrm{cm^{-2} \cdot mag^{-1}}} $ is quite similar to \citet{2018ChAA..42..213L}'s result of  ${2.5\times 10^{21}}$  ${\mathrm{cm^{-2} \cdot mag^{-1}}}$ for the Orion region, calculated by the 21cm spectrum of ${\rm HI}$ and the ${^{12}{\rm CO}}$ emission. This similarity might be attributed to both M17 and Orion being massive star-forming regions, with \citet{2018ChAA..42..213L}'s calculations being based solely on extinction magnitudes of less than 10 mag.
Our result is also similar to \citet{2015MNRAS.448.2187C}'s result of  ${2.41\times 10^{21}}$  ${\mathrm{cm^{-2} \cdot mag^{-1}}}$ for the high galactic latitude interstellar medium, calculated by the 21cm spectrum of ${\rm HI}$ and the ${^{12}{\rm CO}}$ emission, and \citet{2014ApJ...783...17L}'s result of  ${2.7 \times 10^{21}}$  ${\mathrm{cm^{-2} \cdot mag^{-1}}}$ for the all-sky, calculated by the 21cm spectrum of ${\rm HI}$. This similarity might be due to the relatively sparse molecular clouds in low-extinction regions of star-forming areas, causing the distribution of matter to resemble that of the interstellar medium.
The result we obtained for the star-forming regions traced by ${^{12}{\rm CO}}$ is  ${N(\rm H)/A_V= (2.51 \pm 0.19) \times 10^{21}}$  ${\mathrm{cm^{-2} \cdot mag^{-1}}}$, which is higher than \citet{1978ApJ...224..132B}'s the plane of the Milky Way result of  ${1.89 \times 10^{21}}$  ${\mathrm{cm^{-2} \cdot mag^{-1}}}$ obtained by the ${Ly \alpha}$ absorption lines to trace ${\rm HI}$. Moreover, \citet{1978ApJ...224..132B}'s the plane of the Milky Way result of  ${1.89 \times 10^{21}}$  ${\mathrm{cm^{-2} \cdot mag^{-1}}}$ is also higher than \citet{1985ApJ...294..599S}'s all-sky result, which range from  ${1.54 \times 10^{21}}$  ${\mathrm{cm^{-2} \cdot mag^{-1}}}$ to  ${1.67 \times 10^{21}}$  ${\mathrm{cm^{-2} \cdot mag^{-1}}}$, obtained by the ${Ly \alpha}$ absorption lines to trace ${\rm HI}$.
Since there are more stars in the plane of the Milky Way than in the all-sky, and the primary element in stars is hydrogen, it can be inferred that the hydrogen content in the plane of the Milky Way is higher than in the all-sky. Consequently, the ${N(\rm H)/A_V}$ value in the plane of the Milky Way is higher. Star-forming regions, which are areas in the plane of the Milky Way where stars are produced, have even higher hydrogen content. Moreover, in these star-forming regions, hydrogen primarily exists in the form of ${\rm H_{2}}$, with relatively little ${\rm HI}$. Therefore, the ${N(\rm H)/A_V}$ value we calculated by ${\rm H_{2}}$ is higher than the value for the plane of the Milky Way calculated by \citet{1978ApJ...224..132B}, which only considered ${\rm HI}$.
Significant disparities emerge between our result and that of \citet{2012AA...543A.103P} by the 21cm spectrum of ${\rm HI}$ and the ${^{12}{\rm CO}}$ emission, possibly attributable to extragalactic sources situated in high Galactic latitude regions, in contrast to the low Galactic latitude region occupied by M17. These differing spatial positions, coupled with the comparison between average results from multiple extragalactic sources and a single high-density molecular cloud, account for the substantial deviations in our calculated outcomes.

\begin{figure}
	\includegraphics[width=\columnwidth]{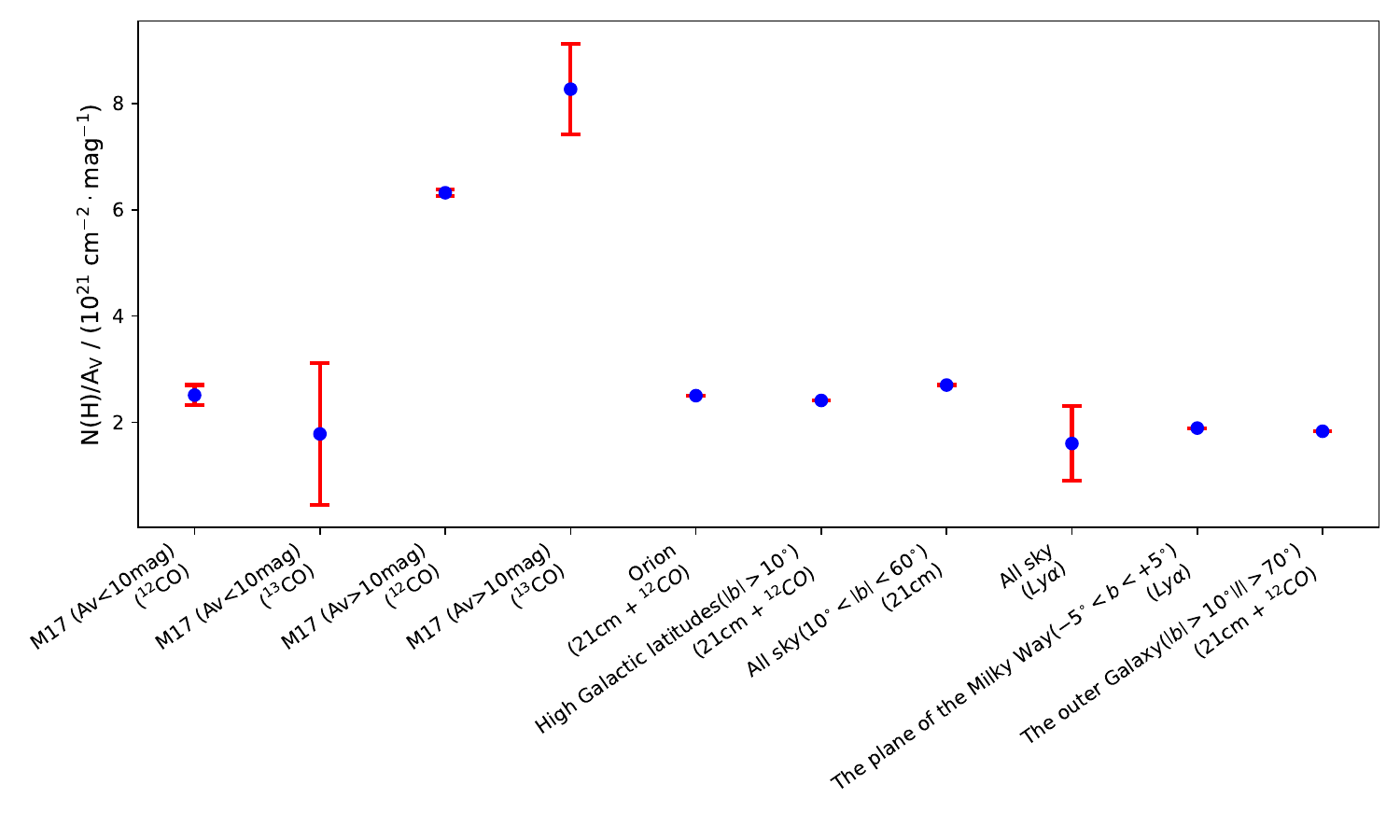}
  \caption{ The comparison of various ${N(\rm H)/A_V}$ values derived from different works. When ${{A_V} \le 10}$ mag, the results of this work are generally consistent with those of previous work. However, when ${{A_V} \ge 10}$ mag, there is a significant difference between the results of this work and previous works. The unit is $\mathrm{10^{21} cm^{-2} \cdot mag^{-1}}$.}
  \label{fig: comparison}
\end{figure}

Table \ref{tab: comparison} and Figure \ref{fig: comparison} presents the comparison of various ${N(\rm H)/A_V}$ values derived from different works. When ${{A_V} \le 10}$ mag, the results of this work are generally consistent with those of previous work, which suggests that our research may be correct. However, when ${{A_V} \ge 10}$ mag, there is a significant difference between the results of this work and previous works, indicating that the GDR in the deep regions of the M17 molecular cloud may be notably different from the other regions. The differences in the GDR may be related to the formation of massive stars in M17, an issue we will explore in detail in Subsection \ref{subsubsec: massive stars}.

\begin{table*}
  \centering
  \caption{comparison with previous work}
  \begin{tabular}{cccc} 
  \hline
  \hline
     Region & tracer of ${N(\rm H)}$ & ${N(\rm H)/A_V}$  $(10^{21}$ ${\mathrm{{cm^{-2}} \cdot mag^{-1})}}$ & Data from sources \\ \hline
     M17 ${(-0.5 ^{\circ} < b < +1.5 ^{\circ })}$ (${{A_V} \le }$ 10 mag) & ${^{12}{\rm CO}}$ &  ${2.51 \pm 0.19}$ & This work \\
     M17 ${(-0.5 ^{\circ} < b < +1.5 ^{\circ })}$ (${{A_V} \le }$ 10 mag) & ${^{13}{\rm CO}}$ &  ${1.78 \pm 1.33}$ & This work \\
     M17 ${(-0.5 ^{\circ} < b < +1.5 ^{\circ })}$ (${{A_V} \ge }$ 10 mag) & ${^{12}{\rm CO}}$ &  ${6.32 \pm 0.06}$ & This work \\
     M17 ${(-0.5 ^{\circ} < b < +1.5 ^{\circ })}$ (${{A_V} \ge }$ 10 mag) & ${^{13}{\rm CO}}$ &  ${8.27 \pm 0.85}$ & This work \\
      Orion ${(-5 ^{\circ} < b < -10 ^{\circ })}$ &  21cm ${+ ^{12}{\rm CO}}$ &  2.5 & \citet{2018ChAA..42..213L} \\  
     High Galactic latitudes ${(\left | b \right | >10 ^{\circ })}$ & 21cm +${^{12}{\rm CO}}$ & {2.41} & \citet{2015MNRAS.448.2187C} \\
     All sky ${(10 ^{\circ} < \left | b \right | < 60 ^{\circ })}$ & 21cm & 2.7 & \citet{2014ApJ...783...17L} \\ 
     All sky & ${Ly \alpha}$ & ${1.54 \sim 1.67}$ & \citet{1985ApJ...294..599S} \\  
     The plane of the Milky Way ${(-5 ^{\circ} < b < +5 ^{\circ })}$& ${Ly \alpha}$ & 1.89 & \citet{1978ApJ...224..132B} \\ 
     The outer Galaxy ${(\left | b \right | >10 ^{\circ } \left | l \right | >70 ^{\circ })}$ & 21cm${ + ^{12}{\rm CO}}$ & 1.83 & \citet{2012AA...543A.103P} \\    
     \hline
  \end{tabular}
  %
  
  \label{tab: comparison}
\end{table*}

This work is based on \citet{2018ChAA..42..213L}, and since M17 is a massive star formation region, we focus on comparing our results with those for Orion in that paper.
The column density of atomic hydrogen is derived from the 21cm spectrum of ${\rm HI}$ of the \emph{Effelsberg-Bonn HI Survey(EBHIS)}. The intensity of ${^{12}{\rm CO}}$ ${J = 1 \to 0}$ line is taken from the \emph{Planck all-sky survey}. The interstellar extinction traced the dust mass using the \emph{2MASS} ${J}$-, ${H}$- and ${K}$-bands photometric database.
Due to the use of different observational data with varying resolutions, they took the resolution of the \emph{Planck} observation as standard to degrade the resolutions of the \emph{EBHIS} and \emph{2MASS} data.
Consequently, the maximum values have been reduced. For instance, the extinction ${A_V} $'s maximum value decreased from 17 mag to 9 mag.
In contrast, our data have a high resolution, with a maximum extinction magnitude of up to 26 mag and maximum hydrogen column densities reaching ${8 \times 10^{23} \mathrm{cm^{-2}}}$ for ${^{13}{\rm CO}}$ and ${2.1 \times 10^{23} \mathrm{cm^{-2}}}$ for ${^{12}{\rm CO}}$. Our observational instruments offer higher sensitivity and better resolution, enabling us to obtain GDRs deep within the molecular cloud.
Table \ref{tab: comparison} provides a comparison. When $A_V \leq 10$ mag, the ${N(\rm H)/A_V}$ values derived from ${^{12}{\rm CO}}$ and ${^{13}{\rm CO}}$ are almost identical, and they are also similar to those for Orion in \citet{2018ChAA..42..213L}. However, the ${N(\rm H)/A_V}$ value derived from ${^{13}{\rm CO}}$ is slightly smaller than that from ${^{12}{\rm CO}}$, possibly due to the dilution of the column density by the small filling factor of ${^{13}{\rm CO}}$ in a thin molecular environment.

\subsection{The effect of massive stars on the GDR} 
\label{subsubsec: massive stars}

Ionizing radiation and stellar winds from massive stars in M17 can significantly impact the physicochemical morphology of the surrounding molecular clouds. This complex process may ultimately affect the GDR. 
Previous studies of massive stars in M17 have provided valuable insights into the nature and abundance of these stars \citep{2008ApJ...686..310H, 2017A&A...604A..78R, 2017yCat..36040078R}.
According to \citet{2015NSLCAS}, we marked the massive stars in M17 on the dust extinction map and the GDR map derived from ${^{13}{\rm CO}}$. 

\begin{figure}
	\includegraphics[width=\columnwidth]{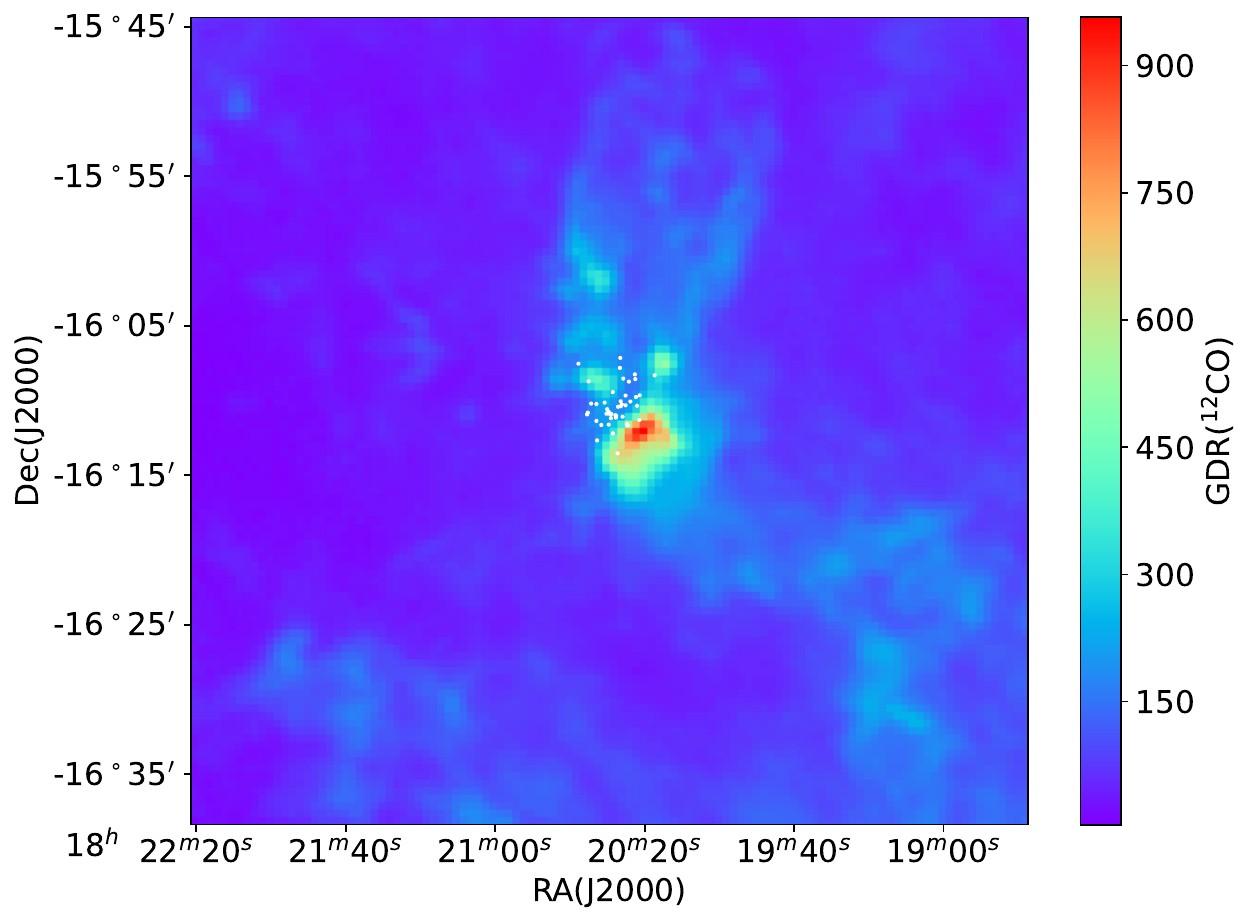}
  \caption{ GDR map derived from ${^{12}{\rm CO}}$ with massive stars marked by white points. These massive stars are distributed in the third-highest GDR region, next to the highest GDR region. And the unit is ${10^{21}}$  ${\mathrm{cm^{-2} \cdot mag^{-1}}}$.}
  \label{fig: massive stars on GDR of 12CO}
\end{figure}

\begin{figure}
	\includegraphics[width=\columnwidth]{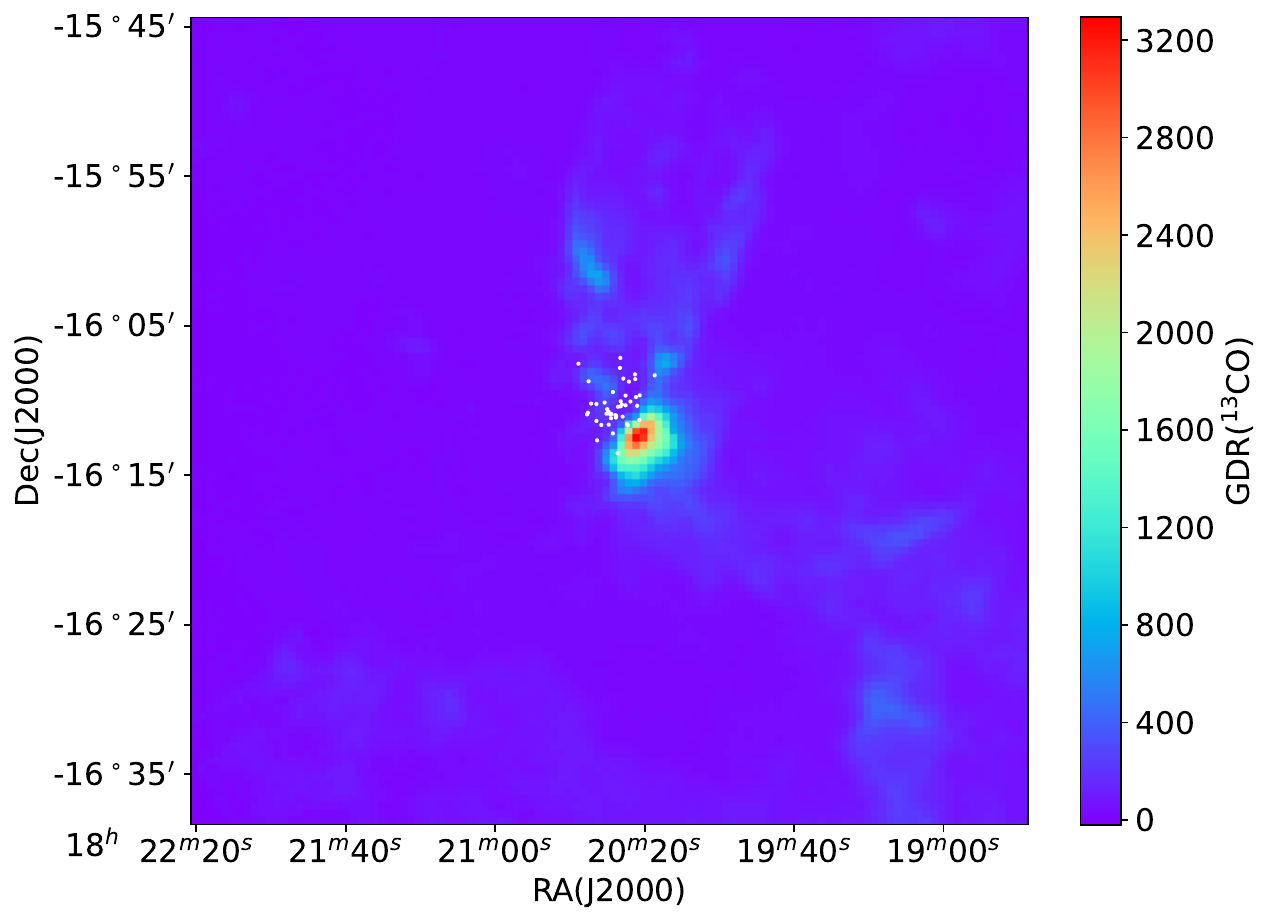}
  \caption{ GDR map derived from ${^{13}{\rm CO}}$ with massive stars marked by white points. And the unit is ${10^{21}}$  ${\mathrm{cm^{-2} \cdot mag^{-1}}}$.}
  \label{fig: massive stars on GDR of 13CO}
\end{figure}

\begin{figure}
	\includegraphics[width=\columnwidth]{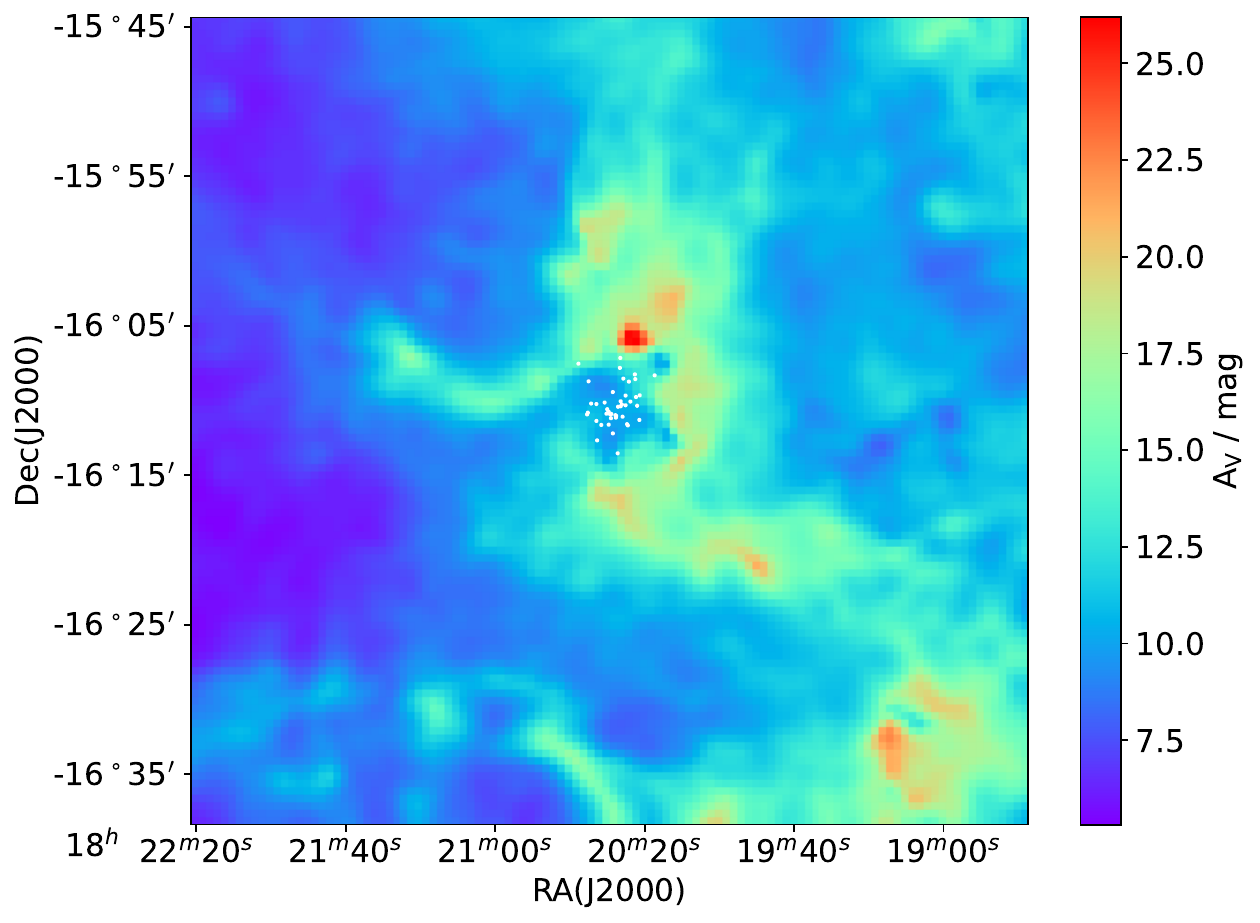}
    \caption{ Extinction map with massive stars marked by white points. These massive stars are distributed within cavities characterized by low extinction magnitude regions surrounded by areas of high extinction magnitude.  And the unit is ${ \mathrm{mag}}$.}
  \label{fig: massive stars on Av}
\end{figure}

Figure \ref{fig: massive stars on GDR of 12CO} depicts the distribution of massive stars in M17 overlaid on the GDR map derived from ${^{12}{\rm CO}}$. 
Figure \ref{fig: massive stars on GDR of 13CO} depicts the distribution of massive stars in M17 overlaid on the GDR map derived from ${^{13}{\rm CO}}$. 
The massive stars are concentrated in the upper left corner of the molecular cloud core, where the GDR values are highest.
Figure \ref{fig: massive stars on Av} depicts the distribution of massive stars in M17 overlaid on the dust extinction map. 
The massive stars are concentrated in the cavity at the centre of the extinction map, where the extinction values are low.
As shown in Figure \ref{fig: massive stars on GDR of 12CO}, \ref{fig: massive stars on GDR of 13CO} and \ref{fig: massive stars on Av}, the region with the highest GDR value (representing the densest molecular cloud nuclei) is situated to the right of the ionized hydrogen region (massive star-forming regions). This phenomenon suggests that the dust extinction magnitude in this area may have been severely underestimated. 
While the ${A_V}$ values in massive star-forming regions are relatively low, the GDR value is not. The GDR values here are slightly lower than that in the densest molecular cloud cores and higher than in other areas. 

The relatively low extinction and high GDR observed near the massive stars in M17 can be attributed to several factors, which we discuss below in terms of both gas and dust properties.

For the gas mass estimates, there are several factors may contribute to the deviations of the hydrogen column density (${N(\rm H)}$), such as the ${\rm CO}$ abundance, ${X_{\rm CO}}$ and ${T_{ex}}$ deviations, the neglected ${\rm H I}$ Mass, the impact of stellar age on ${\rm CO}$ emission. 
(1) ${\rm CO}$ Abundance near the massive stars may be higher than other regions: ${\rm CO}$ freezes on dust grains at low temperatures, and as the temperature increases, ${\rm CO}$ gradually release and exists in the gas phase. The higher temperatures near massive stars may promote the release of ${\rm CO}$, resulting in higher ${\rm CO}$ abundances in these regions compared to others. Additionally, as ${\rm CO}$ release from the dust grains, it no longer participates in the formation of other ions (e.g., ${\rm HCO^{+}}$ and ${\rm N_{2}H^{+}}$) on the dust surface, leading to less depletion and higher ${\rm CO}$ abundances here \citep{2022A&A...662A..39E}. Extinction can also affect the ${\rm CO}$ abundance. When ${A_V \ge 10}$ mag, the gas in the Photo-Dissociation Regions (PDRs) will primarily exist in molecular form (e.g., ${\rm H_{2}}$, ${\rm CO}$) rather than being photo dissociated \citep{1999RvMP...71..173H}. As shown in Figure \ref{fig: massive stars on Av}, the extinction near massive stars in M17 is approximately 11 mag, so the ${\rm CO}$ abundance in this region may be higher than in other areas. The higher ${\rm CO}$ abundance could finally result in an overestimate of the hydrogen column density (${N(\rm H)}$).
(2) The variation of ${X_{\rm CO}}$ and deviations of ${T_{ex}}$ from the assumed relations may affect the estimation of ${N(\rm H)}$. In this work, we  simply referenced the values of ${X_{\rm CO}}$ and ${T_{ex}}$ as given in literature \citet{2013ARA&A..51..207B} and \citet{MolecularAstrophysics}. Factors such as irradiation, turbulence, shocks, and magnetic fields are all elements that affect ${X_{\rm CO}}$ and ${T_{ex}}$. The influence of these factors can lead to biases when estimating ${N(\rm H)}$. In this work, we didn't consider the biases introduced by these factors.
(3) When estimating the mass of ${\rm H}$, we did not consider ${\rm HI}$, which may lead to an underestimation of the H mass \citep{2018ChAA..42..213L}.
(4) Stellar age may also influence ${\rm CO}$ emission. \citet{2019A&A...625A..49K} noted that accretion in young stars is a necessary but insufficient condition for ${\rm CO}$ emission. \citet{2023A&A...676A.122P} found that ${\rm CO}$ emission is more likely to be observed in objects with higher accretion rates and at earlier stages of formation, particularly in M17 Pre-Main Sequence stars where enhanced ${\rm CO}$ emission has been observed. Therefore, the age of the stars is related to ${\rm CO}$ emission and is an important factor influencing ${\rm CO}$ abundances.
The follow-up mid/far infrared observation of interstellar dust \citep{2009ASPC..414..453D}, the observation of ${\rm CO}$ ions \citep{2022A&A...662A..39E} and the follow-up observations of ${\rm HI}$ \citep{2018ChAA..42..213L} may help us to distinguish these possibilities above.

Several factors could introduce biases in the dust mass estimation, like the dust dispersion by massive stars, dust destruction or growth, dust composition, intrinsic extinction and the influence of scattered light on the reddening curve.
(1) The total mass of dust relative to gas could be reduced due to dust dispersion by massive stars. \citet{2024ApJ...974..136S} found that when the stellar mass exceeds 2\Msun, the radiation pressure from the star can disperse the surrounding dust, leading to a decrease in the dust mass and a lower extinction. Additionally, \citet{2023IAUS..362..268K} conducted a detailed model study, revealing the significant effects of stellar winds, radiation pressure, and ultraviolet radiation on the morphology of ${\rm H II}$ regions and their infrared emission. In particular, stellar winds were shown to be the key mechanism for clearing all dust grains from ${\rm H II}$ regions, reducing the dust mass and causing lower extinction. These phenomena could be further studied through numerical simulations in the future.
(2) The destruction of dust may lead to smaller dust particles, and the size of dust particles can have an impact on extinction. The particle collisions in interstellar shocks, sputtering in ${\rm H II}$ regions, and photodesorption by UV radiation are common mechanisms that can lead to the destruction of dust \citep{1979ApJ...231..438D}. Subsequent studies can employ numerical simulations to simulate the evolution of dust in the interstellar medium and analyze the impacts on different destruction mechanisms. In addition, dust can also grow up in dense interstellar medium \citep{2009ASPC..414..453D}. Since dust absorbs the light from stars and re-radiates it in the infrared band, we can understand the dust growth with the dust's temperature, composition, and size distribution by observing the dust spectrum \citep{2009ASPC..414..453D}.
(3) Dust composition may also affect the GDR values. The WD01 model is based on the Milky Way extinction curve and represents a mixture of graphite and silicates in a certain ratio \citep{2001ApJ...548..296W}. However, the composition of dust in M17 may differ from the dust models of the Milky Way. The GDR of the Milky Way is 100 \citep{1978ppim.book.....S}. However, the average GDR with the WD01 model for M17 is higher, which is 221. Moreover, the average GDR derived from graphite in M17 is significantly higher at 945, compared to the average GDR of silicate, which is 122. This phenomenon suggests that the graphite content in M17 might be greater than that in the WD01 model. The discrepancy could be due to the high temperatures generated by massive stars, which can vaporize silicates, leading to an increased proportion of graphite \citep{1979ApJ...231..438D}; alternatively, since silicate grains are generally larger than graphite grains, they might be more easily pushed away by radiation pressure, leaving a higher abundance of graphite behind \citep{2017MNRAS.469..630A}.
The follow up infrared observation of graphite and silicate in M17 \citep{1984ApJ...285...89D} may help us to distinguish these possibilities.
(4) The intrinsic extinction of stars represents an essential component of the total extinction budget. Therefore, failing to account for intrinsic extinction could potentially introduce systematic biases in the derived extinction measurements. Nonetheless, the magnitude of this impact is expected to be comparatively small \citep{2023A&A...676A.122P}.
(5) Scattered light influences the reddening curve, affecting dust extinction measurements. Scattered light exhibits polarization properties, unlike direct stellar radiation. Consequently, polarimetric observations can be employed to account for and mitigate the effects of scattered light, thereby enhancing the precision of dust extinction estimates \citep{2009A&A...493..385K}.

\section{summary }
\label{sec: summary}

Taking advantage of the \emph{UKIRT} and the \emph{MWISP}, we investigated the GDR of the massive star-forming region in M17. We calculated the gas mass of M17 using two tracers, ${^{12}{\rm CO}}$ and ${^{13}{\rm CO}}$ respectively and determined the dust mass from infrared extinction magnitude. The main findings are as follows: 
(1) Based on ${A_V \le}$ 10 mag and ${A_V \ge}$ 10 mag, ${W({\rm CO})/A_V}$ was determined from the fitting in Figure \ref{fig: W1213Av}, and ${{N(\rm H)/A_V}}$ was determined from the fitting in Figure \ref{fig: GDR1213};
(2) Employing the WD01 model, the GDR values are as follows: for ${A_V \leq 10}$ mag, the GDR for ${}^{12}\mathrm{CO}$ is measured at ${118 \pm 9}$, while for ${}^{13}\mathrm{CO}$, it is ${83 \pm 62}$. In contrast, for ${A_V \geq 10}$, the GDR for ${}^{12}\mathrm{CO}$ increases to ${296 \pm 3}$, and for ${}^{13}\mathrm{CO}$, it is ${387 \pm 40}$, and the results for graphite and silicate are presented in Table \ref{tab: GDR};
(3) When $A_V \le 10$ mag, the ${N(\rm H)/A_V}$ values determined in this work are in agreement with those from previous studies. However, when $A_V \ge 10$ mag, the ${N(\rm H)/A_V}$ values determined in this work are significantly higher than the corresponding values from previous studies, as shown in Table \ref{tab: comparison} and Figure \ref{fig: comparison}, indicating that the GDR in the depths of the molecular cloud may be determined;
(4) By marking the massive stars in M17 on the GDR and extinction maps, we found that there are higher GDR values and lower extinction values in these areas compared to the rest of the region in M17. We discussed the potential reasons for this phenomenon and how other observations could distinguish these possibilities.
Especially, the dust composition here seems to contain a higher fraction of graphite than in the WD01 model.

\section*{Acknowledgements}

This project is supported by the National Natural Science Foundation of China (Grant Nos. 12090041, 12090040, 2021YFA1600401, 2021YFA1600400, 12073051).We gratefully acknowledge the staff members of the Qinghai Radio Observing Station at Delingha for their support of the observations. 
Also we gratefully acknowledge Dr Chen Zhiwei for providing the near-infrared images of M17, obtained from the UKIRT telescope.

\section*{Data Availability}

The PSF photometric star catalogue we processed has been uploaded to the database, which is available in [NADC], at https://nadc.china-vo.org/res/r101444/.



\bibliographystyle{mnras}
\bibliography{example} 








\bsp	
\label{lastpage}
\end{document}